\documentclass[reprint,amsmath,amssymb,
 aps,prb,superscriptaddress]{revtex4-2}
\usepackage[utf8]{inputenc}
\usepackage{braket}
\usepackage{physics}
\usepackage{graphicx}
\usepackage{mathtools}
\usepackage{cuted}
\usepackage{xcolor}
\usepackage{enumitem}   
\usepackage{bm}% bold math
\usepackage[hidelinks]{hyperref}% add hypertext capabilities
\usepackage[mathlines]{lineno}% Enable numbering of text and display math
\usepackage{tikz}
\usepackage{standalone}
\usetikzlibrary{fit}
\usepackage[normalem]{ulem}
\definecolor{tensorblue}{rgb}{0.8,0.8,1}
\definecolor{copper}{rgb}{0.72, 0.45, 0.2}
\tikzstyle{tensor}=[rectangle, draw=black, fill=tensorblue, thick, minimum size = 6mm]
\newcommand{\diagram}[1]{ \begin{array}{cc}\begin{tikzpicture}[scale=1] #1 \end{tikzpicture} \end{array} }
\newcommand{\gaurav}[1]{{\color{copper} #1}}
\newcommand{\michael}[1]{{\color{blue} #1}}

\newcommand{\eric}[1]{{\color{cyan} #1}}
\newcommand{\comment}[1]{}
\date{\today}
\begin{document}
\preprint{APS/123-QED}

%\title{Revealing Quantum Phase Diagrams Using Classical Shadows}
%\title{Microcanonical phase diagrams from classical shadows of quantum dynamics}
%\title{Machine learning microcanonical phase diagrams from \michael{shadow tomography} of quantum dynamics}
\title{Revealing microcanonical phases and phase transitions of strongly correlated electrons via time-averaged classical shadows}
\author{Gaurav Gyawali}
\thanks{These authors contributed equally.}
\affiliation{Department of Physics, Cornell University, Ithaca, NY 14853}
\affiliation{Department of Physics, Harvard University, Cambridge, MA 02138, USA}
\author{Mabrur Ahmed}
\thanks{These authors contributed equally.}
\affiliation{Department of Physics, Applied Physics, and Astronomy, Binghamton University, Binghamton, NY 13902}
\author{Eric W. Aspling}
\affiliation{Department of Physics, Applied Physics, and Astronomy, Binghamton University, Binghamton, NY 13902}
\author{Luke Ellert-Beck}
\affiliation{School of Physics and Applied Physics, Southern Illinois University, Carbondale, IL 62901}
\affiliation{Department of Physics, University of Rhode Island, Kingston, RI 02881}
\author{Michael J. Lawler}
\affiliation{Department of Physics, Applied Physics, and Astronomy, Binghamton University, Binghamton, NY 13902}
\affiliation{Department of Physics, Cornell University, Ithaca, NY 14853}
\affiliation{Department of Physics, Harvard University, Cambridge, MA 02138, USA}

\begin{abstract}

  Quantum computers and simulators promise to enable the study of strongly correlated quantum systems. Yet, surprisingly, it is hard for them to compute ground states. They can, however, efficiently compute the dynamics of closed quantum systems. We propose a method to study the quantum thermodynamics of strongly correlated electrons from quantum dynamics. We define time-averaged classical shadows (TACS) and prove it is a classical shadow(CS) of the von Neumann ensemble, the time-averaged density matrix. We then show that the diffusion maps, an unsupervised machine learning algorithm, can efficiently learn the phase diagram and phase transition of the one-dimensional transverse field Ising model both for ground states using CS \emph{and state trajectories} using TACS. It does so from state trajectories by learning features that appear to be susceptibility and entropy from a total of 90,000 shots taken along a path in the microcanonical phase diagram. Our results suggest a low number of shots from quantum simulators can produce quantum thermodynamic data with a quantum advantage.
\end{abstract}
\maketitle

\section{Introduction}

Simulation of strongly correlated electrons in the context of quantum chemistry and condensed matter physics is one of the potential areas in which quantum computers will have a significant advantage over their classical counterparts \cite{Cao_2019, Wecker_2015, Ma_2020}. 
%write a sentence on what the paragraph is about
Strong interaction between the ostensibly simple electrons can give rise to novel phases, including high-temperature superconductivity \cite{highTc_1, highTc_2}, strange metallic behavior\cite{mcgreevy_2010}, fractional excitations \cite{Laughlin_1999}, and quantum spin liquids\cite{Savary_2016}. Condensed matter physics aims to understand these novel behaviors by studying their phase diagrams and phase-defining features. However, failure of perturbation theory and exponential scaling of the Hilbert space for strongly correlated electrons presents a formidable challenge to classical simulation methods such as exact diagonalization, density matrix renormalization group(DMRG) \cite{dmrg1, dmrg2, dmrg3}, quantum Monte-Carlo \cite{qmc} and dynamical mean-field theory\cite{dmft1}. Whereas, this same challenge provides an exciting opportunity for near-term quantum computers.

\comment{2. Opportunities and challenges for quantum computers. Need to change the organization of this paragraph?
Feynman\cite{Feynman1982} conjectured in 1982 that a quantum computer will be able to efficiently simulate quantum mechanical systems more efficiently than a classical computer. The power of a quantum computer lies in its control over the exponentially scaling Hilbert space of qubits, thus turning the curse of dimensionality into a blessing. }

Harnessing the power of a quantum computer to simulate quantum systems \cite{Feynman1982} requires (i) algorithms that can be executed in a reasonable time and (ii) the ability to learn from quantum experiments without exponentially many measurements. Studying the phases via ground state preparation is a QMA-complete problem \cite{kitaev2002classical,Gorman_2021, Gharibian2019complexity, Bookatz2014QMA, kemp2004complexity}, which cannot be carried out in a reasonable time, even with quantum resources. However, performing dynamics on a quantum state is known to be a BQP-hard problem \cite{Baez2020dynamical, Rudi2020approximating}, possible within polynomial time. Likewise, it has been shown that shadow tomography\cite{Aaronson2018} methods such as classical shadows(CS)\cite{huang2020predicting, huang2021provably, Huang2022Learning} are effective at predicting properties using very few measurements. Thus, if we could combine dynamics simulations and classical shadows, we would have an efficient algorithm to simulate condensed matter.

We need to prepare low-energy initial states to use dynamics to simulate condensed matter and exploit quantum ergodicity \cite{von2010proof}. Although preparing ground states of local Hamiltonians on a physical lattice is a challenging problem on a quantum device, it is always possible to prepare some low-energy state with a constant-depth circuit \cite{Ahranov_2013,AnshuNLTS}. Ergodicity then provides a link between statistical averages and time averages obtained from the dynamics of the low-energy state. It is important that the observables of interest, such as the order parameter, equilibrate before the qubits decohere. Nevertheless, rapid equilibration for most local observables is a feature shared by many interacting quantum systems\cite{Wilming_towards_rapid_eq_2017,Malabara_rapid_eq,Hetterich2015}. Thus, equilibrium dynamics of low-energy states appears to be a promising route to studying equilibrium quantum phases and phase transitions.
 
\begin{figure*}[t]
\includegraphics[scale=0.145]{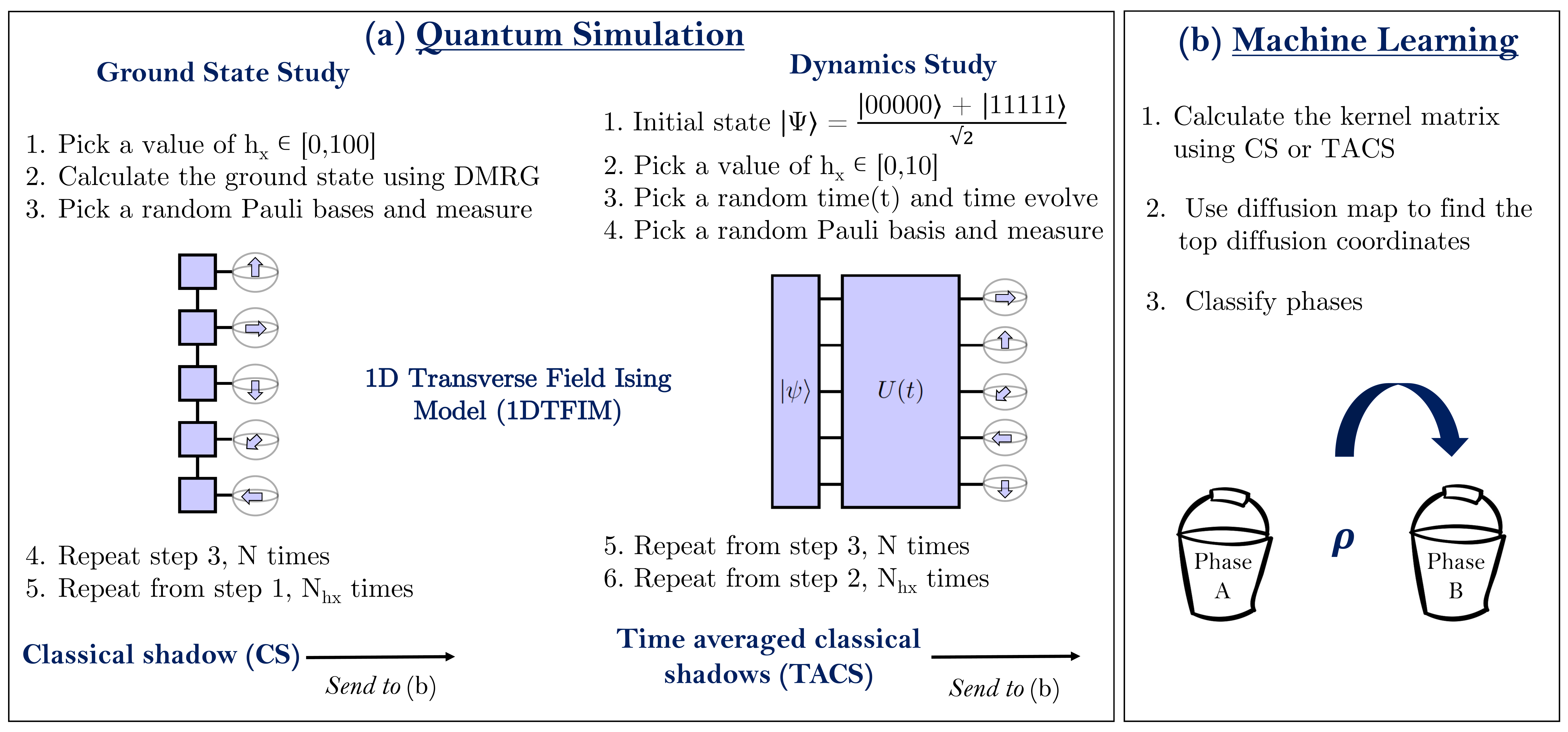}
\caption{Schematic overview of our study. (a) Classical shadows(CS) of ground states  and time-averaged classical shadows(TACS) from dynamics of a time-reversal invariant GHZ state are generated using quantum simulation. (b) The classical data from quantum simulation is then fed into diffusion map, an unsupervised machine learning algorithm to learn the phases.}
\label{Fig:Schematic_diagram}
\end{figure*}

In this manuscript, we present an algorithm to identifying phase diagrams and phase transitions of strongly correlated systems \emph{motivated by how physical quantum systems operate}. It consists of i) identifying an initial state, ii) generating state trajectories by evolving this state in time, iii) using  shadow tomography to convert the quantum state to classical data, and iv) applying unsupervised machine learning methods to discover phases of matter and their phase transitions. A schematic overview of our approach is shown in Fig. \ref{Fig:Schematic_diagram}. 

%We illustrate it with the 1-dimensional Transverse Field Ising Model(1DTFIM). 

%We obtain numerical results using diffusion maps~\cite{coifman2006diffusion, de2008introduction}, an unsupervised machine learning(UL) algorithm,  to machine learn phase features from unlabeled phase diagram data generated by a simulated quantum computer. Specifically, we apply diffusion maps to CS of ground states from a 100 qubit simulation and show they successfully learn the magnetic phase transition of 1-dimensional Transverse Field Ising Model(1DTFIM) (see section II). It does so by \emph{discovering}  the relevant parameters such as magnetization and transverse magnetic field and their \emph{geometric} relationship to each other. These results follow similar machine-learning-from-CS success stories in other contexts\cite{ huang2020predicting,  huang2021provably, randomized_toolbox_2022, Huang2022Learning}. We then show (section III) that with 500 shots per data point, diffusion maps can also identify the phase transition and quantum critical region \emph{when trained on dynamics data from just 20 qubits}. This data is in the form of a representation we introduce: time-averaged classical shadows (TACS)---a shadow tomographic\cite{Aaronson2018} representation of the von Neumann ensemble obtained from the time-averaged density matrix\cite{von2010proof}. \gaurav{The diffusion map identifies the phases and phase transition by learning features from TACS that appear to be susceptibility and entropy.} Our results thus validate the exciting possibility of studying the phases and phase transitions of strongly correlated electrons using quantum computers.

We obtain numerical results using diffusion maps~\cite{coifman2006diffusion, de2008introduction}, an unsupervised machine learning(UL) algorithm to learn phase features from unlabeled data. First, we benchmark diffusion maps on CS data from ground states of a 100-qubit one-dimensional Transverse Field Ising Model(1DTFIM) simulation. It identifies the magnetic phase transition, its continuous nature, and the magnetization behavior as a function of the magnetic field (see section II)---another machine-learning-from-CS success story\cite{ huang2020predicting,  huang2021provably, randomized_toolbox_2022, Huang2022Learning}. Generalizing CS to time-averaged CS (TACS), a shadow tomographic\cite{Aaronson2018} representation of the time-averaged density matrix\cite{von2010proof}, we then show, in section III, on a 20 qubit 1DTFIM, diffusion maps also identify the quantum critical region and cross-overs along a path in the microcanonical phase diagram from a total of 90,000 shots on state trajectories. Diffusion maps do so efficiently by learning features from TACS that appear to be susceptibility and entropy. Hence, we can efficiently study the phases and phase transitions of strongly correlated electrons by quantum-simulating state trajectories.

\section{Von Neumann's Microcanonical Ensemble}
% We should begin by discussing quantum statistical mechanics as a whole and how it is simplier than classical statistical mechanics because the connection between time averages and statistical averages holds even at finite T and finite system sizes. To reach statistical ensembles defined in classical physics, we do need T->infinity and L -> infinite but these steps are not necessary to map time averages to statistical averages.
%\michael{ Surprisingly, for finite system size, this ensemble is not as Boltzmann found for classical particles, equally weighted across phase space, but it is dependent on the initial state which is never forgotten in quantum mechanics.}

A central goal of quantum computing is to build qubits that are completely isolated from their environment. While this is not the case today, the current development of quantum error correction techniques\cite{Knill_2005,nielsen_chuang_2010} suggests it is in our future. Simulating quantum systems on a quantum computer will therefore take place \emph{within the microcanonical ensemble}. But quantum microcanonical dynamics, the evolution of a closed quantum system under Schr\"odinger's equation, does not directly produce the microcanonical ensemble.

Following von Neumann's 1929 paper\cite{von2010proof,goldstein2010normal} on the quantum ergodic theorem, it is straightforward to derive a link between time averages and statistical averages using density matrices. Assuming we start from an initial state $|\psi(0)\rangle$ and evolve under a Hamiltonian $H$ via a quantum circuit algorithm to $|\psi(t)\rangle$, the equilibrium distribution is captured by the von Neumann ensemble, the time average of the density matrix
\begin{equation}\label{eq:rhoT}
    \rho_{vn}%\rho_{vn}
    \! =\! \frac{1}{T}\int_0^T\!\! dt |\psi(t)\rangle\langle\psi(t)| \!\underset{T\to\infty}{\xrightarrow{\hspace*{0.8cm}}}\! \omega\! =\! \sum_n P_n|\psi(0)\rangle\langle\psi(0)|P_n,
\end{equation}
where $P_n$ is a projector onto the $n^\text{th}$ degenerate subspace of the energy eigenvalues i.e. $P_n = \sum_{k\in n}|E_k\rangle\langle E_k|$. The $T\to\infty$ limit, obtained in exponential time\cite{von2010proof,Gogolin_2016, Short_2012,Hetterich2015}, results in equilibration of all observables.
\comment{
We can understand this expression from the snapshot picture of statistical mechanics. An ensemble of states/systems can be thought of as equivalently a history of one system where we record the state of the system in a series of snapshots over time. If the frequency of snapshots is sufficiently low, then each snapshot will be statistically independent of the previous one leading to a probability of $dt/T$ for each snapshot and the above time-averaged expression for the density matrix of this ensemble. }
Existence of $\rho_{vn}$ results in the ergodic principle that time averages of observables should  be captured by the statistical averaging with respect to $\rho_{vn}$.  Specifically, in the Schr\"odinger picture,
\begin{align}
  \langle O\rangle_T &=  \frac{1}{T}\int_0^T dt
    \langle\psi(t)|O|\psi(t)\rangle \\
    &= \frac{1}{T}\int_0^T dt \text{Tr}\left( |\psi(t)\rangle\langle\psi(t)|O\right) \\
    &= \text{Tr} \left(\rho_{vn} O\right) \underset{T \to \infty}{\longrightarrow} \Tr(\omega O)
\end{align}
where again the $T\to\infty$ limit produces equilibration, though here a presumably easier state to reach for it is just necessary for $\rho_{vn}$ and $\omega$ to be indistinguishable to $O$\cite{Hetterich2015}. Thus, the time-averaged density matrix is a link between time averages and statistical averages governed by the von Neumann ensemble $\rho_{vn}$, a link that holds regardless of whether the system equilibrates. 
%\sout{The challenge is to reconcile this with statistical mechanics. How is it that $\rho$ behaves like a microcanonical ensemble of equal a priori probabilities for all accessible states such as}.

The connection to Boltzmann's microcanonical ensemble, obtained by quantizing the classical microcanonical ensemble, is achieved by taking the thermodynamic limit, measuring only coarse-grained observables, requiring non-degenerate energy level spacings/gaps, and considering ``typical'' initial states. The coarse-grained observables used by von Neumann were a commuting set of generators of global symmetries, a restriction more recently generalized\cite{yunger2016microcanonical} as part of the development of quantum thermodynamic resource theories\cite{vinjanampathy2016quantum}.  The typical initial states were the first recorded use of typicality arguments\cite{goldstein2010normal}. Under these circumstances, von Neumann obtained
\begin{equation}
    \omega \sim \rho_{mc} = \frac{1}{\Omega}\sum_{E < E_n < E + \Delta} |E_n\rangle\langle E_n|.
\end{equation}
Namely, for the purposes of computing $\text{Tr}( O\omega)$, there is no difference between using $\omega$ and $\rho_{mc}$, a maximally mixed state within an energy window $[E,E+\Delta]$ containing $\Omega$ states. Von Neumann extended this claim to a second ensemble, $\rho'_{vn} = |\psi(T)\rangle\langle\psi(T)|$, that also satisfies $\rho'_{vn}\sim\rho_{mc}$.  The equivalence to $\rho_{mc}$ is also readily proven with the seemingly stronger requirement that each eigenstate satisfies the eigenstate thermalization hypothesis\cite{deutsch1991quantum,srednicki1994chaos, gong2022bounds}. 
So, in this way, $\rho_{vn}$ reproduces $\rho_{mc}$.

%in the limit of long times and large systems sizes, provided it results from typical initial conditions and we measure coarse-grained observables. \michael{However, for any finite $T$, system size, and observable, $\rho_{vn}$, the ensemble produced by the Schrödinger equation, provides a purely quantum justification of the microcanonical ensemble in the form of the indistinguishability between it and $\rho_{mc}$ to coarse grained observables.}

%Von Neumann's QET sets the conditions for macroscopic equivalence between $\omega$ and $\rho_{mc}$ in the thermodynamic limit. 

%\sout{proves in his theorem that in the thermodynamic limit of $T\to\infty$, $V\to\infty$, $\rho \to \rho_{mc}$ and the microcanonical ensemble emerges, forgetting the initial state $|\psi(0)\rangle$.}

A central new ingredient in von Neumann's approach to describing the microcanonical ensemble is the initial state, which is never fully forgotten in quantum mechanics. 
%In a numerical solution to the above equilibrium description, two things are of utmost importance---reaching the long-time, thermodynamic limit, and understanding the errors that may result along the way. Based on the above logic, an equilibrium phase diagram with finite size errors associated with a reasonable choice of initial conditions seems achievable. 
For each initial condition, it's necessary to check whether the time series was run long enough for relevant observables to reach equilibrium 
%(not all observables equilibrate until an exponential amount of time has passed\cite{Gogolin_2016, Short_2012})
. The same observable will equilibrate at different times depending on the initial conditions. 
%If so, the only errors remaining are of finite size and will be different for each choice of initial conditions. A family of initial conditions that all yield the same microcanonical ensemble, when scaled up, is, therefore, a key resource to study equilibrium statistical mechanics on a quantum computer. Although the equilibrium properties are independent of the initial state, t
It turns out, this time 
%required to reach equilibration depends on the initial state via 
depends on the effective dimension given by $d_\text{eff} = 1/\sum p_k^2$, where $p_k =|\bra{\psi} \ket{E_k}|^2$\cite{Gogolin_2016, Short_2012}. Here, the overlaps $p_k$ measure how many energy eigenstates have a significant weight in $|\psi\rangle$. Remarkably, a large effective dimension results in rapid equilibration. Furthermore, there is a bound on the equilibration time given by the second largest $p_k$ if it is significantly smaller than $1/d_\text{eff}$\cite{Gogolin_2016}. The distribution of the $p_k$'s likely also affects equilibration times\cite{margolus2021counting}. These arguments suggest we choose initial conditions that exhibit a small overlap with most energy levels and or a macroscopic occupation of a single energy level. 
%In practice, we are concerned with the distinguishability of the time-averaged state vs. the equilibrium state in terms of experimentally relevant observables. Therefore, we need not worry about the equilibration of the full-density matrix as long as the equilibration of the relevant order parameters can be achieved.

Because we will use a machine learning method as part of our study of von Neumann's microcanonical ensemble, we need to compare it to what we already know to validate the approach. In the next section, section \ref{sec:groundstatedata}, we will turn to a CS data-driven ground state before continuing to our TACS data-driven thermodynamic study in section \ref{sec:thermo}. 

\section{Ground state data}\label{sec:groundstatedata}
To verify our approach to phase classification and phase-defining feature identification, we first apply it to ground states of the ferromagnetic 1DTFIM defined by the Hamiltvon neuonian:
\begin{equation}
    H_{\text{1DTFIM}} = -\sum_{\langle i, j \rangle} Z_i Z_j + h_x\sum_{i} X_i,
\end{equation}
where $\langle.\rangle$ denotes nearest neighbors, $Z_i$ is the Pauli$-Z$ operator, and $h_x$ is a parameter proportional to the transverse magnetic field. At zero temperature, this model has a ferromagnetic phase for $\abs{h_x}<1$, and a paramagnetic phase for $\abs{h_x}>1$. The ground state study uses diffusion maps to detect the second-order phase transition at $h_x = 1$ for a $100$ site 1DTFIM. The ground states were generated using density matrix renormalization group (DMRG) \cite{dmrg1, dmrg2, dmrg3} with ITensor package\cite{itensor}. 
Training datasets were generated using two kinds of measurements on these ground states-  (i) computational basis measurements to obtain the Z dataset and (ii) measurements on a random Pauli basis to obtain the CS dataset.

\subsection{Classical Shadows}

\begin{figure*}[ht]
\begin{center}
\includegraphics[scale=0.9]{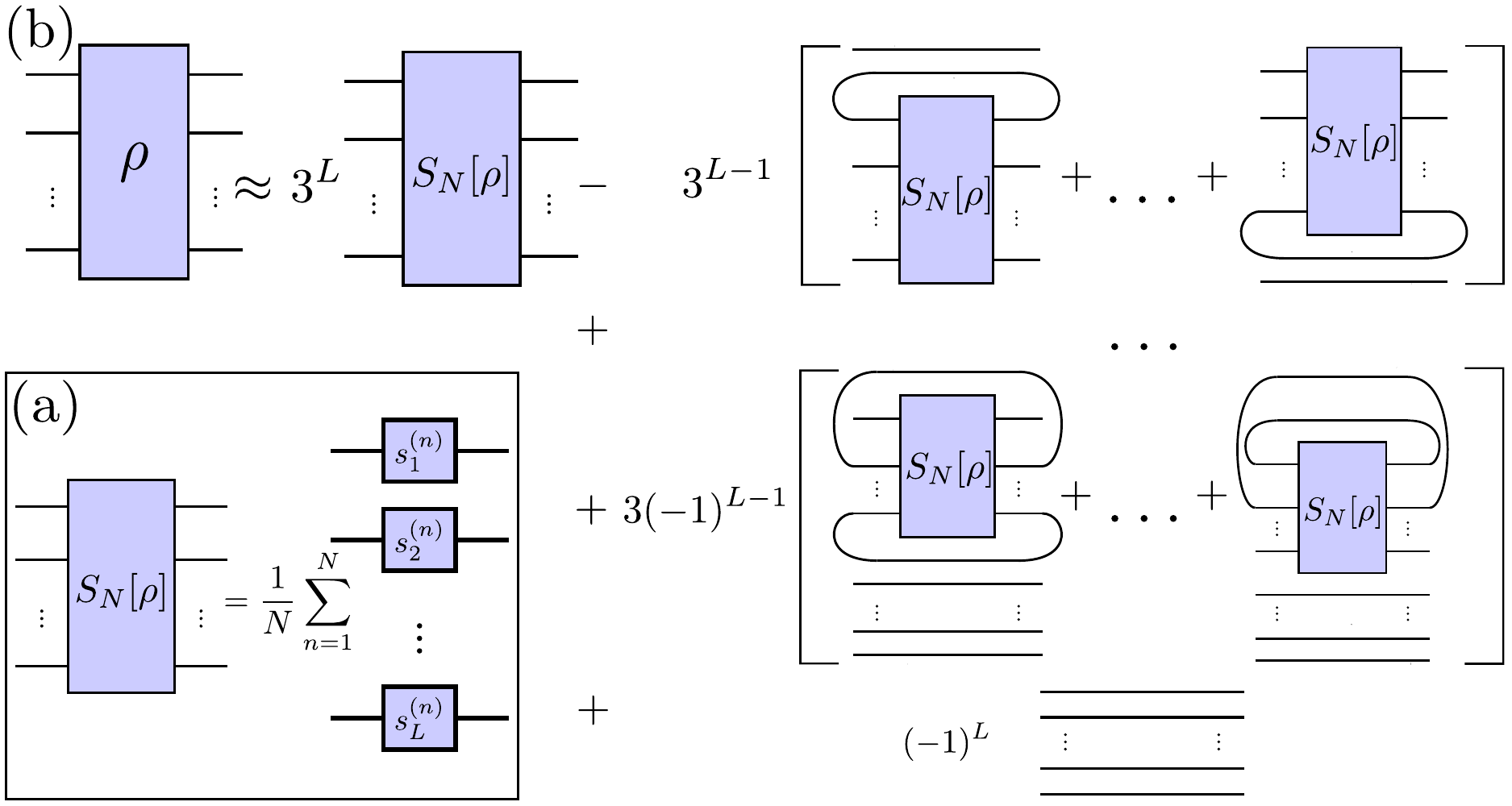}
\caption{Diagrammatic description of classical shadows showing a linear relationship between $S_N[\rho]$ and estimator $\sigma_N[\rho]\approx\rho$. (a) The full-density matrix $\rho$ can be approximated by summing over reduced classical shadows with a coefficient that grows exponentially in the number of remaining qubits. (b) A classical shadow $S_N[\rho]$ is obtained by summing over $N$ measurement outcomes on random Pauli bases. For a given $N$, a reduced density matrix, that involves smaller coefficients in the expansion, can be approximated more accurately compared to the full density matrix.}
\label{Fig:CS_diagramatic}
\end{center}
\end{figure*}

\comment{\michael{In principle, the choice of shadow tomography impacts what observables are visible in the data. Similarly, in finite-time microcanonical dynamics simulations, coarse grained observables equilibrate while fine grained ones do not. So, to study quantum dynamics on a quantum computer, \sout{we should} \eric{one must} adjust the choice of shadow tomography to best observe the behavior of equilibrating observables.}}

Obtaining any useful information from a quantum computer requires performing measurements on a quantum state, which is destructive to the quantum information by nature. Since the dimension of the Hilbert space increases exponentially in the number of qubits, a naive strategy to learn the state requires an exponentially large number of copies. Aaronson\cite{Aaronson2018} introduced an alternative method using the notion of shadow tomography, an approximate classical description of the quantum state, in which $M$ properties of a quantum state can be estimated with error $\epsilon$ by only $O(\frac{M}{\epsilon^2})$ copies of the state. We can think of a shadow as an approximation of a quantum state $\rho$ by summing over measurement outcomes $x$, obtained by performing measurements on bases $b$ for a quantum state $x$, i.e.
%\begin{equation}\label{eq:cs}
%    \rho = \sum_{b,x}P(b)
%        \mathcal{P}_{b,x}
%        |\psi_0\rangle\langle\psi_0|
%        \mathcal{P}_{b,x},
%\end{equation}
\begin{equation}\label{eq:cs}
    S[\rho] = \sum_{b,x}P(b)
        \mathcal{P}_{b,x}
        \rho
        \mathcal{P}_{b,x},
\end{equation}
where $ \mathcal{P}_{b,x}$ is a projector onto the measurement outcome $x$ on basis $b$, and $P(b)$ is the probability of choosing $b$.

Based on this notion, Huang et al.\cite{Huang2019predicting, huang2021provably, Huang2022Learning} developed an algorithm called classical shadows and showed that it is highly successful at learning the properties of a many-body system. Two kinds of measurement protocols were proposed to construct classical shadows-  (i) random Clifford measurements on the entire Hilbert space; (ii) random single-qubit Pauli measurements. Protocol (ii) results in very shallow measurement circuits and thus is more suitable for the NISQ-era\cite{PreskillNISQ} hardware. \comment{With random Pauli measurements, they show that predicting linear functions with error $\epsilon$ requires at least $\text{log}(M) \text{max}_i \text{tr}(O_i^2)/\epsilon^2$ copies of the state.} After measuring each of the qubits in some random Pauli basis $X$, $Y$ or $Z$ with outcomes $\pm1$, the post-measurement wavefunction is given by the product state $\left|s^{(n)}\right\rangle=\bigotimes_{l=1}^{L}\left|s_{l}^{(n)}\right\rangle$. Here, $\left|s_{l}^{(n)}\right\rangle \in\{|0\rangle,|1\rangle,|+\rangle,|-\rangle,|\mathrm{i}+\rangle,|\mathrm{i}-\rangle\}$ is a Pauli basis state to which the $l^{th}$ qubit has collapsed. The classical shadow $S_N[\rho]$ is obtained by summing over $N$ such randomized measurement outcomes as follows (also see Fig. \ref{Fig:CS_diagramatic}(b))
\begin{align}
    S_N[\rho] &= \frac{1}{N} \sum_{n=1}^N \ket{s^{(n)}} \bra{s^{(n)}}\\
    &= \frac{1}{N} \sum_{n=1}^N \ket{s_1^{(n)}} \bra{s_1^{(n)}} \otimes \cdots \otimes  \ket{s_L^{(n)}}\bra{s_L^{(n)}}.
\end{align}

The underlying quantum state $\rho$ can be approximated by adding the reduced classical shadows (see Fig. \ref{Fig:CS_diagramatic}(a)).  This sum simplifies to the following expression from Ref. \cite{Huang2019predicting, huang2021provably, Huang2022Learning}
\begin{equation}\label{eq:csshots}
    \rho \approx \sigma_{N}(\rho)=\frac{1}{N} \sum_{n=1}^{N} \sigma_{1}^{(n)} \otimes \cdots \otimes \sigma_{L}^{(n)},
\end{equation}
where
\begin{equation}
\sigma_{l}^{(n)}=3\left|s_{l}^{(n)}\right\rangle\left\langle s_{l}^{(n)}\right|-\mathbb{I}.
\end{equation}

The definition of $S_N(\rho)$ presented above is different from Refs. \onlinecite{Huang2019predicting, huang2021provably} which defines it to be the dataset of shots itself and not the density matrix obtained from these shots. But both definitions are complete for Fig. \ref{Fig:CS_diagramatic} (whose derivation from tensor network diagrams is presented in Appendix A) shows the density matrix $S_N(\rho)$ defined above is linearly related to the estimator $\sigma_N(\rho)$ of the quantum state $\rho$ obtained by Refs. \onlinecite{Huang2019predicting, huang2021provably, Huang2022Learning}. Hence, the two definitions are informationally equivalent. 

Although estimating the exact density matrix requires $N \rightarrow \infty$, we still desire to predict various linear as well as nonlinear functions of $\rho$ (e.g., $\Tr(O\rho)$ and $\Tr (\rho \text{log}(\rho))$ respectively). This can be achieved with $N \propto \log(L) 4^k/\epsilon^2$ copies of the state,  where $k$ is the locality of operator $O$ \cite{randomized_toolbox_2022}. It was shown in Ref. \cite{huang2021provably} that classical machine learning algorithms can efficiently predict the ground state properties of gapped Hamiltonians in finite spatial dimensions after learning the classical shadows from a training set. An example of interest is classifying the quantum phases of matter. Classifying the symmetry-breaking phases is conceptually simple because it involves calculating $\text{tr}(\rho O)$ for some k-local observable $O$, such that $\text{tr}(\rho O) \geq 1 \: \forall \: \rho \in \text{ phase A}$ and  $\text{tr}(\rho O) \leq -1 \: \forall \: \rho \in \text{ phase B}$.

In contrast to classifying symmetry-breaking phases, capturing continuous phase transitions and classifying topological phases involves nonlinear-in-$\rho$ observables like critical exponents and entropy, which are harder to estimate than linear observables. Learning such nonlinear functions requires an expressive ML model. A central object in kernel-based ML is the kernel function, a local similarity measure in the feature space where the samples live. Ref. \cite{huang2021provably} proposed a kernel based on mapping from classical shadows to a high-dimensional feature space that includes the polynomial expansion of many-body reduced density matrices. Learning nonlinear functions requires access to k-body reduced density matrices, where k may be large, but with enough shots, classical shadows can accomplish this.Using such a kernel, Ref. \onlinecite{huang2021provably} found a rigorous guarantee that a classical ML algorithm can efficiently classify phases of matter, including the topological phases. We will employ this kernel to study the continuous phases transition in the 1DTFIM.

%Thus, it is beneficial\michael{[vague]} to devise a mapping from classical shadows to a high-dimensional feature space that includes the polynomial expansion of many-body reduced density matrices\michael{[it sounds here like this is an obvious way to construct a kernel for this problem]} and encapsulate this feature space in a kernel function. Using such kernels then establishes a rigorous guarantee\michael{[why? Is all that matters is whether the kernel is non-linear in rho?]} that a classical ML model can efficiently classify nonlinear functions, phase transitions, and even topological phases of matter\cite{huang2021provably}. \michael{Hence, we will employ such a kernel to study the continuous phase transition in the TFIM.}

%\section{A discussion of Ground States using generated data}
\subsection{Machine Learning Method: Diffusion Maps}
\label{sec:ML}
\comment{Classical shadows provide an approximate information about the full state of a quantum computer where the accuracy of the approximation depends just on the number of shots[ref]. But rather than work directly with this approximation, such as by computing observables that are accurately captured by the data, we will apply machine learning (ML) methods whose success depends on this approximation. As we will see, these ML methods can still identify macroscopic parameters of interest, such as the z-magnetization and $h_{x}$ values of the quantum states, as well as identify phases and phase transitions, even though the data we feed it (the classical shadows) contain far less information than the full description of the quantum states.}

%\subsection{Transverse Field Ising Model}

% mabrur
Let's now turn to the final step in our approach: applying an unsupervised machine learning method called diffusion maps \cite{coifman2006diffusion, de2008introduction} to extract features from the shadow tomography data. A diffusion map is a nonlinear dimensionality reduction technique that relies on learning the underlying manifold from which the data points have been generated. Recently, this method was used to identify phases in systems with complex order parameters, which are difficult to learn using linear methods (such as principal component analysis (PCA)\cite{carleo2019machine}). Examples of such phase identification studies include: topological phases and phase transitions \cite{rodriguez2019identifying}, incommensurate phases, and many-body localized phases in quantum systems \cite{lidiak2020unsupervised}.

In the application of diffusion maps, we imagine a random walk on a dataset  $X\left(x_{1}, x_{2},\ldots, x_{N}\right)$, where the $x_{n}$ are estimators $\sigma_N(\rho)$ of density matrices $\rho$ obtained from different points in the phase diagram. The transition probability $P(j\backslash i)$ of jumping from $x_{i}$ to $x_{j}$ in a single ``timestep'' is proportional to the kernel function $k(x_i,x_j)$, a non-negative similarity measure between the two data points. Here we use the classical shadow kernel function prescribed in~\cite{huang2021provably}, defined to be for two points $x$ and $\tilde x$ in the dataset
\begin{equation}\label{eq2}
  \begin{aligned}[b]
    & k^{(shadow)}(x,\tilde x) = k^{(shadow)}(\sigma_N(\rho),\sigma_N(\tilde\rho))  {}\\ 
    &= \exp\!\left(\!\sum_{n,n'=1}^N\!\frac{\tau}{N^2}\!     \exp\!\left(\!\frac{\gamma}{L}\!\sum_{l=1}^L\! \Tr[\sigma_{l}^{(n)} \tilde\sigma_{l}^{(n')}]\!\right)\! \right).
  \end{aligned}
\end{equation}
This kernel measures the local similarity between  $x$ and $\tilde x$ by comparing the  trace distance between the CS estimates of all k-reduced density matrices. For the diagonal components ($x=\tilde x$) of the kernel matrix, the trace distance between the k-reduced density matrices is the $2^{nd}$ Renyi-entropy. We then construct a transition probability matrix $P$ such that 
\begin{equation}
\label{eqn:P-matrix}
   {P(j\backslash i)%P_{ij} = p(i,j)}
   } 
   = \frac{k^{(shadow)}(i,j)}{\sum_l k^{(shadow)}(i,l)}.
\end{equation}
%where $d_{X}$ is the normalization constant: $d(j) = \sum_{i=1}^N k(i,j)$. 
%I changed this noting that $\sum_j P(j\backslash i) = 1$ is a conditional probability of going to $j$ given the starting point $i$ (michael). 
After $t$ timesteps of the random walk, the transition probabilities are given by the matrix $P^t$, where $P_{ij}^t$ gives the probability of going from $x_{i}$ to $x_{j}$ in $t$ timesteps, it's a sum of the probabilities associated with all of the possible paths to go from $x_{i}$ to $x_{j}$ in $t$ timesteps. As $t$ increases, the diffusion process unfolds, where data points situated along the overall geometric structure of the dataset become more strongly connected because of the abundance of strongly connected intermediate points along the way. 

Given this random walk, we can define a `diffusion distance' to quantify this idea of connectivity between two data points:
\begin{equation}\label{eq:dmdistance}
    D_{t}(x_{i},x_{j})^2 = \sum_{m=1}^N|P_{im}^t-P_{mj}^t|^2,
\end{equation}
where the bigger the diffusion distance, the weaker the connection between them. This allows us to map the data points onto a new `diffusion space' so that the diffusion distance in data space is equal to the Euclidean distance in this new space. Following Ref. \onlinecite{coifman2006diffusion}, we will do so with the map:
\begin{equation}\label{eq5}
    x_{i}\rightarrow y_{i} = \begin{bmatrix}
       \lambda_{1}^t\psi_{1}(i),  &  
       \lambda_{2}^t\psi_{2}(i), & 
       \cdots, & 
       \lambda_{N-1}^t\psi_{N-1}(i)
    \end{bmatrix}
\end{equation}
where $\lambda_{k}$ and $\psi_{k}$ are eigenvalues and right eigenvectors of the matrix $P^t$, $\psi_{k}(i)$ is the $i$-the element of the $k$-th eigenvector. Then the diffusion distance is,
\begin{equation}\label{eq6}
    D_{t}^2(x_{i},x_{j}) = |y_{i} - y_{j}|^2 \\
        = \sum_{k=1}^{N-1} (\lambda_{k}^{t})^2\left[\psi_{k}(i) - \psi_{k}(j)\right]^2. 
\end{equation}
Plotting data in this new space provides an intuitive geometric picture of the data manifold.  

The map provides several features we can exploit when interpreting the data. In equations (\ref{eq5}) and (\ref{eq6}) the $k=0$ component is ignored because the leading eigenvector, $\psi_{0}(i) = \frac{1}{\sqrt{N}},\; \lambda_{0}=1$, is constant for all $i$ by the Perron-Frobenius theorem. But this constant eigenvector impacts the other eigenvectors: We note  add a constant term $\psi_k(i)\to\psi_k(i) + C$ to all {other eigenvectors and still preserve the diffusion distance. Additionally, equation (\ref{eq6}) suggests a dimensionality reduction, as the terms with bigger $\lambda_{k}$ will dominate the sum increasingly as $t\rightarrow\infty$. So plotting the data $x_i$ in the truncated space $[\lambda_1^t\psi_1(i),\lambda_2^t\psi_2(i),\lambda_M^t\ldots,\psi_M(i)$ with $M$ determined by keeping only the significant eigenvalues $\lambda_m\gg \lambda_{M+1}$, implies they are accurately separated by distance $D_t(x_i,x_j)$. Lastly, we see that $t$ is arbitrary. Choosing different $t$ rescales the lengths of each component of the vector $y$. Hence, the data exists on a hyperplane in Euclidean space up to a certain shift in the origin and a one-parameter rescaling of the axes. 

So, using the properties of the diffusion space, we can define \emph{diffusion coordinates} $\text{dc1}(i) = A(\psi_1(i)+C), \text{dc2}(i)=B(\psi_2(i)+C),\ldots$ that map the data $x_i$ onto a Euclidean space that through the choice of constants $A,B,C,\ldots$ \emph{allow us to interpret the coordinates of each point and visualize the geometry of the data}.

%So only the first $m$ diffusion coordinates associated with the bigger eigenvalues ($m << N$) are kept. The data points are then projected onto this m-dimensional space to reveal features in the dataset.

\section{Phase Classification of Ground States}
For the ground state study, we used two datasets, one the computational basis measurements and the other generated via CS tomography. The first (Z-dataset) contains qubit measurements only along the Z-axis [in Eq. \ref{eq:cs}, $P(b)=1$ for $b=Z$, $P(b)=0$ for all other b]. While the other (CS-dataset) has randomized Pauli measurements using the CS method [$P(b)=\frac{1}{3}$ for $b \in \{X, Y, Z\}$]. Both of them contain 200 100-spin 1DTFIM ground state shots for each state obtained from different $h_{x}$ values ($h_{x}$ ranging from 0.1 to 100).  Since the $Z$-magnetization is the order parameter for TFIM, the UL algorithm should be able to learn the phases of the model from the Z-dataset. Using this knowledge, we compare the Z-dataset and the CS-data to see if the UL algorithm can successfully identify phases in each case and if so how it does so.

%$Z$ is the relevant observable to identify phases in the TFIM model, we want to find out how well the machine can learn from the CS-dataset compared to the Z-dataset. After using diffusion maps to project the data points onto a lower dimensional space (diffusion space), we used k-means clustering in that space to classify them. Figure (\ref{ground}) shows our results:  

% ground state results begin
\begin{figure}[t]
\centering
\includegraphics[scale=1]{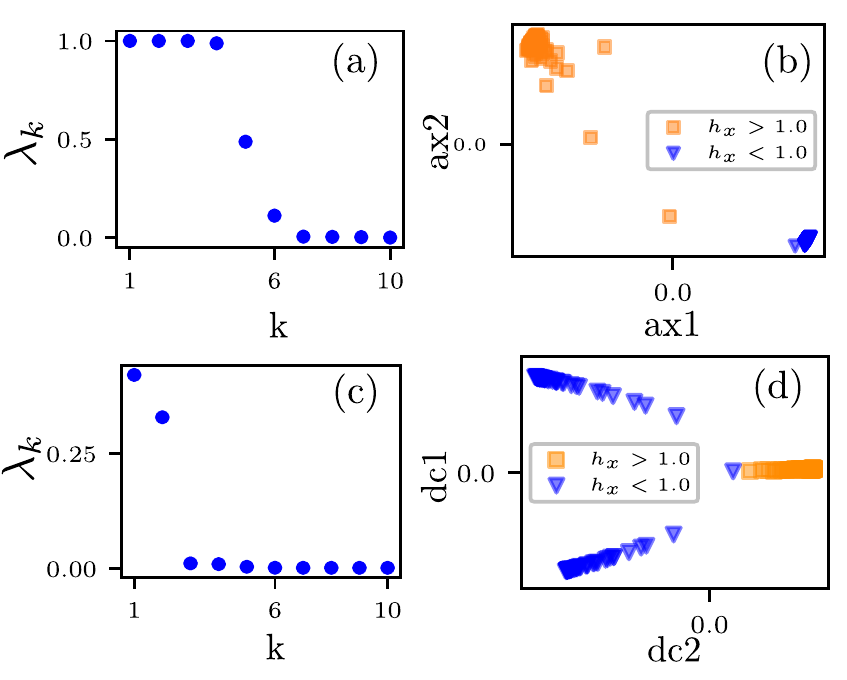}
\caption{Learning phases from ground state (CS and Z) data. (a) and (c) show the 10 largest eigenvalues of the $P$-matrix (excluding the trivial $k=0$) for (a) Z-data and (c) CS-data datasets. (b) Z-data points in 5D diffusion space visualized in 2D, using metric MDS. Clustering clearly emerges based on the two phases of the model. (d) CS-data points in 2D diffusion space, the figure reveals the symmetry-breaking phase transition. In this case, there is a direct correlation of relevant parameters, the z-magnetization and the $h_{x}$-values, with machine-learned diffusion coordinates dc1 and dc2 respectively. Notice the CS-data shows one cluster in (d) with non-trivial geometry associated with the critical point, while the Z-data shows two clusters in (b) with trivial geometry and no understanding of the critical point. 
%which is a consequence of the shadow kernel function for the former is not spin-flip symmetric while that of the latter is.
}
\label{ground}    
\end{figure} 
%ground state results end

By deploying diffusion maps armed with the shadows kernel function (Eq. \ref{eq2}) utilized for both data sets as our UL model, we are able to identify the phases from both the Z data and the CS data. In both cases, we set $\tau=1$, $\gamma=1$ in Eq. \ref{eq2}. For the Z data, we chose the first five non-trivial eigenvectors as the diffusion space basis vectors because the $P$-matrix (Eq. \ref{eqn:P-matrix}) eigenvalue spectrum shows the first five eigenvalues to be larger than others (Fig. \ref{ground}(a)). Mapping the states from Z data onto this five-dimensional diffusion space, we found that clear clustering emerges based on the phases of the states. We used multidimensional scaling (MDS), a dimensionality reduction method \cite{borg2005modern} that seeks to preserve point-to-point distances, to project these states onto a 2D plane. We see a clear separation of the two phases even on this 2D reduced space (Fig. \ref{ground}(b)), indicating the machine's success in identifying the two phases.   

From the CS data, the unsupervised learning algorithm was also able to learn about the phases and the underlying parameters of the model. We can see in Fig. \ref{ground}(c) that the $P$ matrix has two non-trivial eigenvalues larger than the rest. The eigenvectors corresponding to these two eigenvalues are the basis vectors of the reduced diffusion space. Figure (\ref{ground}(d)) shows all of the ground state classical shadows projected onto this two-dimensional plane. It shows three groups- the top left and the bottom left are the all-up and all-down states, whereas the group on the center right are states in the disordered phase. This closely resembles the spontaneous symmetry-breaking phase transition of 1DTFIM\cite{sachdev_2011}. The learned diffusion coordinates (dc1 and dc2) have direct correlations with the magnetization $<M_{z}>$ and the field values $h_{x}$ respectively, as shown in appendix C (Fig. \ref{fig:correlation}).  

The above-mentioned clustering is dependent on the number of snapshots $N$ for a given state. However, the dimensionality reduction and the subsequent clustering will settle down after a minimum value of $N$ has been reached ($N_{c}$). Increasing the value of $N$ beyond that point does not change the results in any significant way. We find $N_{c}$ (Z data) $< N_{c}$ (CS data), so it is easier for the algorithm to learn the phase space structure of the Z-data (hence fewer snapshots are required) than the CS-data.

A striking feature of the diffusion map results presented in Fig. \ref{ground} is the geometry it reveals about the data. In the case of Z-data, it finds the data is separated into two distinct clusters(Fig. \ref{ground}(b)) while in the CS-data case it finds only one cluster but that this cluster has a non-trivial geometry with three curves meeting at the critical point (Fig. \ref{ground}(d)). This geometry is directly a consequence of visualizing of the data space through the lens of the kernel function that defines distances between data points via Eqs. \ref{eqn:P-matrix} and \ref{eq:dmdistance}. Another kernel function might see the same CS-data as separate clusters. Hence, it is a striking feature of the kernel of Eq. \ref{eq2} that it can capture the full geometry of phase defining features in the TFIM model.

%\section{A discussion of generating Equilibrium data from a Quantum Computer}

\section{Microcanonical Dynamics of the 1DTFIM}\label{sec:thermo}
Despite the successful identification of the ground state phases with the CS+ML model, the fact remains that the problem of calculating ground-states is a QMA-hard problem \cite{Bookatz2014QMA}. We now turn to
%While there is no shortage of algorithms designed to generate ground-states, the necessity of 
an algorithm built using Schrödinger dynamics, a known BQP class algorithm, that aims to reveal the microcanonical phase diagram as a proxy for a ground state study. 

Consider now Fig. \ref{fig:phasediagram}, a sketch of the thermodynamic phase diagram of the 1DTFIM relating internal energy $E=\langle\psi(0)|\hat H|\psi(0)\rangle$ to the transverse magnetic field $h_x$ inspired by Ref. \cite{wu2018crossovers}. This phase diagram is relevant for a microcanonical dynamics study governed by the entropy $S(E,h_x)$. It exists even for a simulation over a finite time $T$ and with a finite number of spins $N$ and a specific choice of initial conditions but with finite $T$, finite $N$, and initial choice-dependent errors that round phase transitions.
%that round the phase transitions and render the results dependent on initial conditions. 
We present in this figure our expectations for the phase diagram in this context, pointing out phase transitions where the phase diagram will sharpen in the thermodynamic limit. We further highlight the path through the phase diagram carried out by our simulations below, showing that we expect it to cross the quantum critical region and so be sensitive to the phase diagram at a rounded level even in the long-$T$, large-$N$ limit. 

\begin{figure}
    \centering 
    \includegraphics{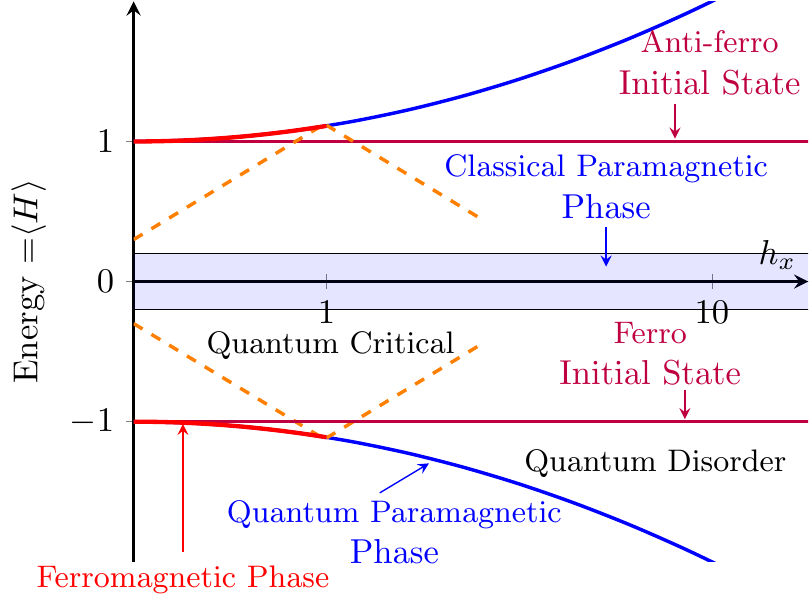}
    \caption{
    A sketch of the expected 1DTFIM phase diagram at finite $T$ and finite $N$ as a function of internal energy $E$ and transverse magnetic field $h_x$. This diagram is a modification of the canonical ensemble representation of the phase diagram in Ref. \onlinecite{wu2018crossovers}, adapted to the microcanonical ensemble. The spectrum is mirror symmetric about $E=0$ due to the chiral symmetry $\mathcal{C}=ZYZYZY...$ \comment{that anticommutes with the Hamiltonian.}}
    \label{fig:phasediagram}
\end{figure}

\comment{\gaurav{\emph{Where is a good place to add this?} A similar algorithm called thermal shadows has been recently proposed in literature\cite{Predicting_Gibbs}. However, this requires a large number of controlled unitaries and optimization for the quantum signal processing\cite{QSP} subroutine. There is also quantum imaginary time evolution(qite) \cite{QITE}, which approximates a k-local imaginary time evolution operator (non-unitary) by C-local unitary where $C>k$. This involves approximating all the non-unitary operators corresponding to each term in the Hamiltonian by using a larger unitary, whose size grows with the correlations(C)  in the system. Our algorithm seems very straightforward compared to this, so will potentially replace qite. There is also another algorithm based on the zeros of partition function \cite{Partition_zeros}.}}

%\subsection{Results}
To reveal the phase diagram expected from microcanonical dynamics presented in Fig. \ref{fig:phasediagram}, we need an experimentally producible classical representation of the quantum data obtained from a microcanonical dynamics simulation. Noticing that the time-averaged integral amounts to an expectation value of the pure state density matrix $|\psi(t)\rangle\langle\psi(t)|$ over the probability distribution $P_T(t) = (1/T)\left(\Theta(T-t\rangle)-\Theta(-t)\right)$, where $\Theta(x)$ is the Heaviside step function, we see we can construct \emph{time-averaged classical shadows}(TACS) by the quantum channel
\begin{equation}\label{eq:tacs}
    TACS\left[\rho \right] = \lim_{T\to\infty}\int dt \sum_{b,\sigma}P_T(t)P(b)
        \mathcal{P}_{b,\sigma}
        |\psi(t)\rangle\langle\psi(t)|
        \mathcal{P}_{b,\sigma}.
\end{equation}
Hence by sampling the joint probability distribution $P_T(t)P(b)$ to obtain $(t_i,b_i)$, $i=1\ldots N$, and then measuring one shot $\sigma_i$ from $|\psi(t_i)\rangle$ in basis $b_i$ we obtain a finite-shot TACS via
\begin{equation}\label{eq:tacsshots}
    TACS_N\left[\rho \right] = \comment{\lim_{T\to\infty}}
    \comment{\lim_{N\to\infty}
    \frac{1}{N}}
    \sum_{i=1}^{N}
    |t_ib_i\sigma_i\rangle\langle t_ib_i\sigma_i|
\end{equation}
This approach captures the power of CS tomography and enables an experimental study of microcanonical thermodynamics.

With this shadow tomography method in mind, we ran quantum dynamics simulations of 1DTFIM using the TDVP algorithm\cite{tdvp,timeevomps} starting from the GHZ state $\ket{\psi(0)} = \frac{\ket{000\cdots} + \ket{111\cdots}}{\sqrt{2}}$ to generate TACS data from 500 randomly sampled dimensionless time values between $t=10.0$ to $t=20.0$ and 187 randomly sampled $h_x$ field values between $h_x = 0.1$ to $h_x = 10.0$. %For each $h_{x}$, a TACS was obtained at random from the $N = 500$ shots times over the fixed $T$ time interval following Eq. \ref{eq:tacsshots}. 
An example code to generate TACS dataset for 1DTFIM is available in our github repository \cite{TACSgithub}. These 187 TACS were the data points with which we performed unsupervised learning by constructing the $187\times187$ kernel matrix using the shadow kernel in equation (\ref{eq2}) and then using diffusion maps for dimensionality reduction.

A key element needed to obtain reasonable results from the above calculation is an initial state that equilibrates within the chosen time window for observables of interest that are accurately captured by the chosen method of shadow tomography. In the above case, we started from a GHZ state because it equilibrated efficiently for local observables, as shown in Appendix \ref{app:equilibration}. Presumably, this equilibration would occur even faster if we broke the integrability of the 1D TFIM model by adding certain additional terms to the Hamiltonian. Hence, up to possibly finite size effects, thermodynamic observables in our results should behave as expected.

Fig. \ref{fig:dynamics} shows our results. The first two eigenvectors were chosen as our diffusion space basis vectors because as Fig. \ref{fig:dynamics}(inset) shows, those are the two dominant non-trivial eigenvalues in the spectrum. The rest of the eigenvectors go to zero as the number of data points increases. Projecting the states onto this two-dimensional diffusion space, we see that the states all fall on a curve in this hyperplane along which the value of $h_{x}$ increases monotonically, and the inflection point neighborhood of the curve coincides with the quantum critical region [see \ref{appendix_crit}]. Therefore, as in the ground state study, it is apparent that the unsupervised learning algorithm was able to infer two phases from the data via a single cluster with non-trivial geometry. Presumably, taking paths through the phase diagram closer to the critical point, we would see the inflection point sharpen, leading to a singular point in the data manifold at the critical point. However, unlike the ground state study, it is not obvious what the two diffusion coordinates dc1 and dc2 correspond to. To identify these, we need to study observables capable of capturing the phase-defining features and see which correlate with these learned coordinates.

%In contrast to our ground state study, we were not able to map the diffusion coordinates to the phase-defining observables.

%dynamics results start
\begin{figure}[t]
\centering
\includegraphics[scale=0.93]{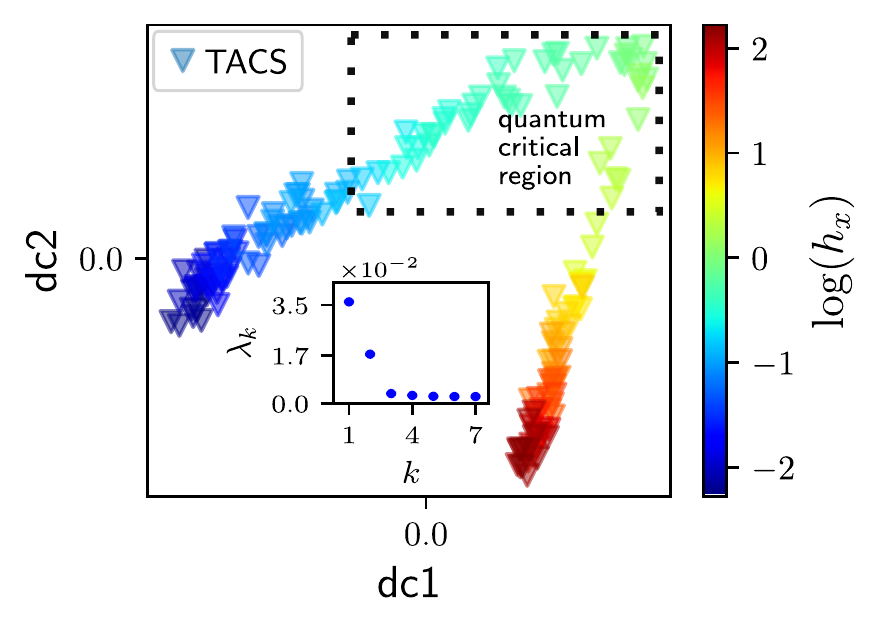}
\caption{Phase identification from dynamics data. The eigenvalue spectrum (inset) obtained from diffusion maps shows the two largest eigenvalues corresponding to the two dominant diffusion coordinates dc1 and dc2 (the trivial point $k=0$, is not shown). The TACS data points largely fall on a two-stranded curve parameterized by $h_{x}$ in this 2D reduced diffusion space. The quantum critical region (in green) coincides with the inflection point neighborhood of the curve, with points on the left strand belonging mostly to the ordered phase while points on the  right strand belong to the disordered phase.}
\label{fig:dynamics}    
\end{figure}
%dynamics results end

% More things to explain including what the order parameter is, and how it is expected to behave through the phase transition
% Explain Multidimensional scaling and the process to look at the zoomed in clusters. 
% Diagrams to show are; dc1 vs dc2, dc1 vs M, correlation function and index for different h_x values, and corr_z vs h_x
% To identify the reliability of the algorithmic clustering, we will zoom into a smaller resolution of h_x values not at the phase transition. 

As a preliminary exploration of phase-defining observables, a straightforward first approach is to check whether the diffusion coordinates obey power laws consistent with the known quantum critical point. In Fig. \ref{fig:renyi}(a), we plot dc2, which diverges as it approaches the critical point with critical exponents $p_- = 0.58 \pm 0.05$ and $p_+ = 0.7 \pm 0.1$. We have shifted the diffusion coordinates by $C = -0.0027$ since the diffusion distance is invariant under an overall shift of the origin as mentioned in Sec. \ref{sec:ML} and this renders it positive. Remarkably, this shift \emph{simultaneously renders both dc1 and dc2 positive}. However, our errors in the exponents are hard to estimate.  Suppose we view the unknown variable $C$ as a Gaussian distribution. In that case, the corresponding distribution of $p_{\pm}$ from our predictions is highly non-Gaussian (see Appendix \ref{app:nu}). The closest known critical exponent is $\nu=1$ \cite{cardy1996scaling} (see also wikipedia\cite{wiki:ising}). However, we found that the observable that qualitatively matches the diverging behavior at $h_x=1$ is xx component of the susceptibility, which we define as:
\begin{equation}
    \chi_{ab} = \frac{1}{L^2} \sum_{ij} \expval{\sigma^a_i \sigma^b_j},
\end{equation}
where $\sigma^a \in \{X,Y,Z\}$ is a $2 \times 2$ Pauli matrix. This is different than the usual definition obtained by summing over the connected correlations. The resemblance between dc2 and susceptibility here is only qualitative, so the critical exponents do not match. Since there are a number of observables, such as $\chi_{xx}, \chi_{yy}$ and $\chi_{zz}$, that are equally likely candidates to define the phases, we conjecture that dc2 could be some combination of these.

This leaves the puzzle of determining dc1, which neither diverges nor shows a power law behavior. When inverted, it appears qualitatively similar to the ZZ component of the susceptibility (see Sec. \ref{sec:susceptibilities}) but we turn to calculate the $2^{nd}$ Renyi entropy given the diagonal components of the kernel function capture this quantity. Entropy is the key thermodynamic potential of the microcanonical ensemble that certainly captures phase transitions and from which all important phase-defining features could be extracted. It is also known that the quantum critical region features the interplay of equilibrium and quantum fluctuations leading to the entropy being maximized \cite{ Wu_2011_entropyaccumulation, wu2018crossovers}.  Renyi entropy is a lower bound for the von Neumann entropy and has the same limits---it vanishes for pure states and reaches $N$ for the maximally mixed state. We do so by first computing the purity using eqn \ref{eq:csshots} as follows
\begin{align}
    \gamma\left[\rho \right] &= \text{Tr}\left[ \rho^2 \right] \\
    &\approx \frac{1}{N^2} \sum_{n\neq n' = 1}^{N} \text{Tr}\left[ \sigma_1^{(n)} \sigma_1^{(n')}\right]\times \cdots \times \text{Tr}\left[ \sigma_L^{(n)} \sigma_L^{(n')}\right],
\end{align}
then the Renyi entropy is $S_2= -\log_2 \gamma$. 

\begin{figure}[t]
\centering
\includegraphics[scale=1.0]{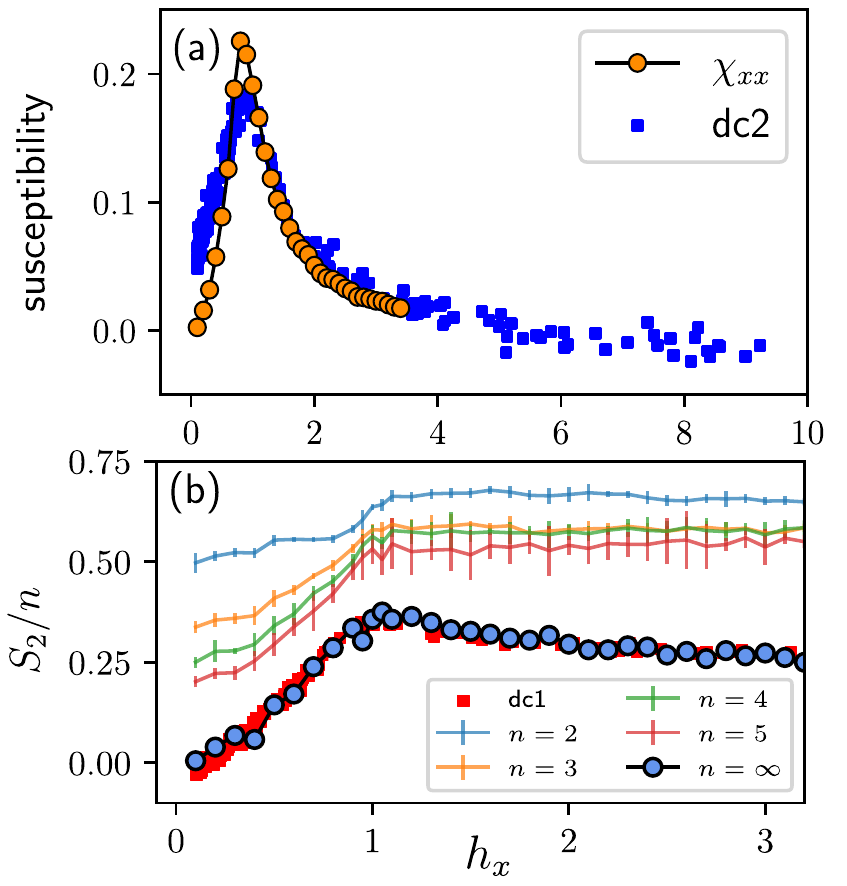}
\caption{Interpretation of dc1 and dc2 for microcanonical dynamics of 1DTFIM.  The divergent behavior of dc2 qualitatively matches the xx component of the susceptibility, computed using 100k shot TACS data for a 10-site 1DTFIM, denoted by orange circles (b) dc1 matches the Bayesian inference estimate for the second Renyi entropy per site ($S_2/n$) in the thermodynamic limit ($n=\infty$). Bayesian inference is performed on $n$-body entropies for $n=1-5$, also computed using the 10-site dataset.}
%\caption{The $2^{nd}$ Renyi entropy per site ($S_2/n$) for a 10-site 1DTFIM computed from TACS consisting of 100k shots. $n=\infty$ shows the Bayesian inference estimation for the thermodynamic limit and infinite TACS shots. The top diffusion coordinate (dc1) shows a similar trend as ($S_2$)  }
\label{fig:renyi}    
\end{figure}

Figure \ref{fig:renyi}(b) presents the Renyi entropy as calculated from TACS shots on a $L=10$ qubit system using exact diagonalization. We use Bayesian inference, as detailed in appendix \ref{app:bi} (see also Ref. \onlinecite{lukens2021bayesian}) to extract the \comment{long-$T$,} large-$n$, large-$N$ predictions with $N=100,000$ shots per $h_x$ value and 35 $h_x$ values between $h_x =0.1$ and $h_x = 3.3$. The range of time sampled for this particular calculation was between $t=5$ and $t=25$. The results show clear evidence that entropy is maximized in the quantum critical region around $h_x=1$ as expected from Fig. 2 of \cite{wu2018crossovers}. Error estimates for these values were obtained in Ref. \onlinecite{Shack}, and given by:
\begin{equation}
    N \geq \frac{4^{n+1}\gamma}{\epsilon^2\delta}
\end{equation}
Where $1-\delta$ is the probability of obtaining a good TACS dataset and $\epsilon$ is the additive error. For a $\delta = 0.33$ (67 percentile), and $N=100,000$ shots, we find an additive error for the $n=5$ entropy curve plotted in Fig. \ref{fig:renyi} at $h_x=1.0$ of $S_2/n \approx 0.5 \pm 0.24$. This is larger than the observed error shown via error bars in Fig. \ref{fig:renyi} but only within a factor of order 1. Hence, by using $3,300,000$ shots, we have estimated the thermodynamic entropy as a function of $h_x$ that reproduces the expected maxima at the critical point. 

Given an estimate of the entropy, we lastly turn to plotting it alongside dc1 to discover that it is highly correlated with this observable. Although dc1 was calculated using only 500 shots per $h_x$ value, orders of magnitude less than the number of shots needed for accurate Bayesian inference estimation, the diffusion process is able to combine information across different $h_x$ values without any supervision.

In summary, the diffusion map was able to learn phase-defining features from TACS and used these features to map the data points\comment{time-averaged density matrices} as a function of the model parameter $h_x$ onto a curve in the two-dimensional plane with geometry that reveals the quantum phase transition. 

\section{Outlook}
In this paper, we have identified an approach to studying quantum thermodynamics on a quantum computer in a way that is suitable for studying quantum materials, their phases, and their phase transitions. This approach consists of 
\begin{enumerate}[label=(\roman*)]
    \item preparing a low-depth initial state for which relevant observables are observed to equilibrate within the coherence time of the quantum computer,
    \item time evolving this state using a quantum algorithm to a randomly chosen time point $t$ within some time interval,
    \item extending shadow tomography methods to obtain a physically useful representation of the von Neumann ensemble %time-averaged density matrix, 
    such as TACS used in this paper, and
    \item employing an unsupervised machine learning method to discover the phase diagram, with kernel methods such as diffusion maps employing well designed kernels showing promise.
\end{enumerate}
Our approach parallels statistical mechanics calculations on classical Hamiltonians, where a random initial state is prepared, a Metropolis Monte-Carlo algorithm is run beginning from this state, and data is collected and analyzed using traditional observables and more recently machine learning methods. Our results, demonstrating the existence of a quantum phase transition and the ability to map out regions of the phase diagram by a careful choice of initial conditions, show promise. 

%\michael{[This last paragraph needs work. It needs to think big, focus on what are the key-resources needed for our approach to work compared to the competition instead of say its simplicity.]}
%\gaurav{@Michael, how does this sound now?}\michael{I guess I don't like the claim that we are simpler than others. How about focusing on a) that our approach requires dynamics simulations out to a time $T$ in which relevant observables equilibrate, b) We take TACS data that requires only a small number of shots, 500 in the present problem, per point in the phase diagram and c) For basic observables, we do not need a large system $L$, finding we could compute a reasonable thermodynamic entropy with just $L=10$ to $L=20$; then with these resources established, compare to the competition and suitability for NISQ devices and quantum simulators?}

There are several resources needed to carry out the microcanonical dynamics simulations. A key resource is a low-energy state that equilibrates within the accessible time scale and can also be prepared easily. For local Hamiltonians on physical lattices, we can always find low-energy states which can be prepared with constant-depth circuits \cite{Ahranov_2013}. Another resource to carry out dynamics simulations to a time $T$ in which the relevant local observables equilibrate ($T=25$ in our simulations). It allows one to exploit advances in variational time evolution algorithms\cite{Yao_2021, Yuan_2019, VFF_2020}, which are especially suitable for the NISQ era due to robustness to noise and ability to go beyond the coherence time of quantum computers. Finally, we need the ability to perform time averages by sampling the state $|\psi(t)\rangle\langle\psi(t)|$ at least at the Nyquist rate determined by the bandwidth, which is linear in the system size. We produced a TACS dataset consisting of 500 shots from the equilibrium dynamics starting from a GHZ state at each of 187 points in the phase diagram; these resources were all that were required for diffusion maps to learn the phases and identify the phase transition for a 20-qubit system. 
%Shannon-Nyquist theorem states the number of samples scales with the inverse bandwidth, thus only polynomially in system size (L). 
Somewhat different resources, a system size of $L=10$, and $N=100,000$ shots at $33$ points in the phase diagram, were required for Bayesian inference to obtain reasonable estimates for the thermodynamic entropy. 

We believe these resources are significantly smaller than those of other proposed methods for studying thermodynamics on quantum computers. For example, the overhead from using ancillas as a heat bath in the existing methods to study thermodynamics on a quantum computer \cite{Predicting_Gibbs, QITE, Partition_zeros} is not an issue with our approach. The resources required also open up an exciting possibility of employing new generation of quantum simulators \cite{Rydberg_atoms,Karamlou2022} to study quantum thermodynamics as they too can simulate quantum dynamics and are capable of performing randomized measurements. Finally, a possible direction for future research would be to identify and benchmark a strongly interacting system with initial states that will yield quantum advantage in the near term.

\begin{acknowledgments}
We thank Mikhail Lukin, Katherine Van Kirk, Nishad Maskara, Yanting Teng, Subir Sachdev, Daniel Parker, and Anurag Anshu for useful discussions. This material is based upon work supported by the National Science Foundation under Grants No. OAC-1940243 and OAC-1940260.
\end{acknowledgments}

\appendix

\section{Diagramatic understanding of classical shadows}
\comment{Performing a large number of Pauli classical shadows measurements depolarizes the density matrix.} In this section, we will develop a diagrammatic understanding of the classical shadows. We will work in the superoperator formalism where the indices of the density matrix $\rho_{ij}$ are grouped together to make a vector $|\rho\rangle\rangle$ and the product $A\rho B$ translates to an operator $ A\otimes B^T$ acting on the vector $|\rho\rangle\rangle$.

In general, we can view the outcomes obtained from many classical shadows measurements on the same prepared state $\rho$ as defining an ensemble of states $S\left[ \rho \right]$ where
\begin{equation}\label{eq:CSmap}
S[\rho] = \sum_{b,x} P(b) \mathcal{P}_{b,x} \rho\mathcal{P}_{b,x} = \bigg[\sum_{b,x} P(b)\mathcal{P}_{b,x}\otimes\mathcal{P}_{b,x}^T\bigg]|\rho\rangle\rangle,
\end{equation}
and $\mathcal{P}_{b,x}$ is the projector in basis $b$ onto qubit state $|x\rangle$. To break this down into manageable parts, let's start from the one qubit case and work our way up to $N$ qubits.

\emph{One qubit case.} In the one qubit case, we generate classical shadows samples for bases $b\in\{X,Y,Z\}$ with uniform probabilities i.e. $P(b)=1/3$. We can thus express the one-qubit version of \ref{eq:CSmap} with the following  diagram:
\begin{align}
	S\left[ \rho \right] = \frac{1}{3} \sum_{b,x} 
	\diagram{
		\node[tensor] at (1,0) (A) {$P_{b x}$};
		\node[tensor] at (2,0) (B) {$\rho$};
		\node[tensor] at (3,0) (C) {$P_{b x}$};
		\draw[-,thick] (A)--(B);
		\draw[-,thick] (B)--(C);
		\draw[-,thick] (0,0)--node[above] {} ++(A) ;
		\draw[-,thick] (C)--node[above] {} ++(1,0);}.
	\end{align}
Using cap and cup notation, we can redraw this in the superoperator form:
\begin{align}
	S[\rho] = \frac{1}{3} \sum_{b,x} 
	\diagram{
		\node[tensor] at (1,0) (A) {$P_{b x}$};
		\node[tensor] at (1,1) (B) {$P_{b x}$};
		\node[tensor] at (2.5,0) (C) {$\rho$};
		\draw[-,thick] (0,2)--(4,2);
		\draw[-,thick] (0,0)--++(A) ;
		\draw[-,thick] (0,1)--++(B) ;
		\draw[-,thick] (A)--(C);
		\draw[-,thick] (B)--++(2.5,0); 
		\draw[-,thick] (C)--++(1,0);
		\draw[thick] (3.5,0) .. controls (3.5+1/4,0) and (3.5+2/4,3/4) .. (3.5,1);
		\draw[thick] (0,1) .. controls (0-1/4,1) and (0-2/4,1+3/4) .. (0,2);
		\draw[thick][red,thick,dashed] (0.25,-0.5) rectangle (1.75,1.5);
	}
	\end{align}
The highlighted box  can be viewed as a superoperator acting on the space of linear operators  $\rho$.  Remarkably, this particular superoperator consisting of a product of two projection operators simplifies substantially i.e.
\begin{align}
	\sum_{b,x} 
	\diagram{
		\node[tensor] at (1,0) (A) {$P_{b x}$};
		\node[tensor] at (1,1) (B) {$P_{b x}$};
		\draw[-,thick] (0,0)--++(A) ;
		\draw[-,thick] (0,1)--++(B) ;
		\draw[-,thick] (A)--++(1,0);
		\draw[-,thick] (B)--++(1,0); 
	} =
	\left[ \;
	\diagram{
		\draw[-,thick] (0,0)--(0,1);
		\draw[-,thick] (0.5,0)--(0.5,1);} 
	+
	\diagram{
		\draw[-,thick] (0,0)--(1,0);
		\draw[-,thick] (0,0.5)--(1,0.5);}
		\;
	\right].
	\end{align}
They amount to the sum of an identity and a cup-cap product. Using this simplification we recognize the one qubit case as the depolarizing map
\begin{align}
	S \left[ \rho \right] = \frac{1}{3}
	\left[ 
	\diagram{
		\node[tensor] at (1,0) (A) {$\rho$};
		\draw[-,thick] (0,0)--(A);
		\draw[-,thick] (A)--(2,0);
	}
	+
	\diagram{\draw[-,thick] (0,0)--(0,0);\draw[-,thick] (0,0.20)--(1,0.20);}
	\right],
	\end{align}
where we further simplified using
\begin{align}
	\diagram{
		\def\c{1/4}
		\def\x{1/2}
		\draw[thick](-\x,0) -- (\x,0);
		\draw[thick](-1/4,3/4) -- (1/4,3/4);
		\draw[thick](-\x,0) .. controls (-\x-\c,0) and (-\x-2*\c,3/4) .. (-1/4,3/4);
		\draw[thick](\x,0) .. controls (\x+\c,0) and (\x+2*\c,3/4) .. (1/4,3/4);	
		\node[tensor] at (0,0) (A) {$\rho$};} = 1
	\end{align}
Inverting, we can extract the original density matrix (full tomography) via

\begin{align}
	\diagram{
		\node[tensor] at (1,0) (A) {$\rho$};
		\draw[-,thick] (0,0)--(A);
		\draw[-,thick] (A)--(2,0);
	} 
	=
	3 \diagram{
		\node[tensor] at (1,0) (A) {$S\left[\rho \right]$};
		\draw[-,thick] (0,0)--(A);
		\draw[-,thick] (A)--(2,0);
	} -
	\diagram{\draw[-,thick] (0,0)--(0,0);\draw[-,thick] (0,0.20)--(1,0.20);}
	%drawing a null to establish origin
	\end{align}
Going from diagrammatic results to equations gives 
\begin{equation}
S \left[ \rho \right] = \frac{1}{3}(\rho + I) \, \implies \, \rho = 3S \left[\rho \right] - I,
\end{equation}
which is a single-qubit depolarizing channel.  

%using $\mathcal{P}_{X,0} = (I+X)/2$, $\mathcal{P}_{X,1} = (I-X)/2$, and similarly for the $Y$ and $Z$ bases, we obtain
%\begin{equation}
%    CS(\hat\rho) = \frac{1}{3}(\frac{3}{2}I\otimes I + \frac{1}{2}\left(X\otimes X-Y\otimes Y+Z\otimes Z\right)|\rho\rangle\rangle
%\end{equation}
%where the minus sign for the $Y\otimes Y$-term arises due to the transpose in Eq. \ref{eq:CSmap}. In the computational $Z$-basis, these terms can be explicitly computed, summed and we arrive 
%\begin{equation}
%    CS(\hat\rho) \rightarrow{Z} \frac{1}{3}\begin{pmatrix} 
%    2\rho_{00} + \rho_{11} & \rho_{01}\\
%    \rho_{10} & 2\rho_{00}+\rho_{11}
%    \end{pmatrix}
%\end{equation}
%Then using $\rho_{00}+\rho_{11}=1$ we obtain finally
%\begin{equation}
%    CS(\hat\rho) = (\hat I+\hat\rho)/3
%\end{equation}
%In this way, we see the one-qubit case amounts to a depolarization of the density matrix. 

\emph{Two-qubit case.} In the two-qubit case, the CS map of Eq. \ref{eq:CSmap} takes the form 
\begin{align}
\frac{1}{9}\sum _{b, x} \sum_{b', x'}
\diagram{
	\node[tensor] at (1,0) (A) {$P_{b' x'}$};
	\node[tensor] at (1,1) (B) {$P_{bx}$};
	\draw[-,thick] (0,0)--++(A) ;
	\draw[-,thick] (0,1)--++(B) ;
	\draw[-,thick] (A)--++(1,0);
	\draw[-,thick] (B)--++(1,0); 
	\draw[fill=tensorblue, thick] (2,-0.5) rectangle (3,1.5) (C)  node[pos=0.5] {$\rho$};
	\draw[-,thick]  (3,0)--++(1,0);
	\draw[-,thick]  (3,1)--++(1,0);
	\node[tensor] at (4,0) (C) {$P_{b' x'}$};
	\node[tensor] at (4,1) (D) {$P_{bx}$};
	\draw[-,thick] (C)--++(1,0);
	\draw[-,thick] (D)--++(1,0); 
}. 
\end{align}
Again, we can add cups and caps to express it in superoperator form:
\begin{align}
\diagram{
	\node[tensor] at (1,0) (A) {$P_{b' x'}$};
	\node[tensor] at (1,1) (B) {$P_{bx}$};
	\node[tensor] at (1,-1) (C) {$P_{b'x'}$};
	\node[tensor] at (1,2) (D) {$P_{bx}$};
	\draw[-,thick] (0,3)--++(4,0);
	\draw[-,thick] (0,-2)--++(4,0);
	\draw[-,thick] (0,0)--++(A);
	\draw[-,thick] (0,1)--++(B);
	\draw[-,thick] (0,-1)--++(C);
	\draw[-,thick] (0,2)--++(D);
	\draw[-,thick] (A)--++(1,0);
	\draw[-,thick] (B)--++(1,0); 
	\draw[-,thick] (C)--++(3,0);
	\draw[-,thick] (D)--++(3,0); 
	\draw[fill=tensorblue, thick] (2,-0.5) rectangle (3,1.5) (C)  node[pos=0.5] {$\rho$};
	\draw[-,thick]  (3,0)--++(1,0);
	\draw[-,thick]  (3,1)--++(1,0);
	\draw[thick](4,1) .. controls (4+1/4,1) and (4+2/4,3/4+1) .. (4,2);
	\draw[thick](4,0) .. controls (4+1/4,0) and (4+2/4,-3/4) .. (4,-1);
	\draw[thick](0,2) .. controls (-1/4,2) and (-2/4,3/4+2) .. (0,3);
	\draw[thick](0,-1) .. controls (-1/4,-1) and (-2/4,-3/4-1) .. (0,-2);
	\draw[thick][red,thick,dashed] (0.2,0.5) rectangle (1.8,2.5);
}. 
\end{align}
This diagram shows that the two-qubit projection operators produce the same structure on each qubit as they did in the one-qubit case. Applying the same simplification as before,  we arrive at 

\begin{align}
\frac{1}{9} 
&\left[
\diagram{
	\draw[fill=tensorblue, thick] (0,-0.5) rectangle (1,1.5) (C)  node[pos=0.5] {$\rho$};
	\draw[-,thick]  (1,0)--++(0.5,0);
	\draw[-,thick]  (1,1)--++(0.5,0);
	\draw[-,thick]  (-0.5,0)--++(0.5,0);
	\draw[-,thick]  (-0.5,1)--++(0.5,0);
}  +
\diagram{
	\draw[fill=tensorblue, thick] (0,-0.5) rectangle (1,1.5) (C)  node[pos=0.5] {$\rho$};
	\draw[-,thick]  (1,0)--++(0.5,0);
	\draw[-,thick]  (1,1)--++(0.5,0);
	\draw[-,thick]  (-0.5,0)--++(0.5,0);
	\draw[-,thick]  (-0.5,1)--++(0.5,0);
	\draw[-,thick]  (-0.5,1.7)--++(2,0);
	\draw[-,thick]  (-0.5,2)--++(2,0);
	\draw[thick](-0.5,1) .. controls (-0.5-1/8,1) and (-0.5-2/4,3/6+1) .. (-0.5,1.7);
	\draw[thick](1.5,1) .. controls (1.5+1/8,1) and (1.5+2/8,3/6+1) .. (1.5,1.7);
} \right.\nonumber\\
& \left.+ \diagram{
	\draw[fill=tensorblue, thick] (0,-0.5) rectangle (1,1.5) (C)  node[pos=0.5] {$\rho$};
	\draw[-,thick]  (1,0)--++(0.5,0);
	\draw[-,thick]  (1,1)--++(0.5,0);
	\draw[-,thick]  (-0.5,0)--++(0.5,0);
	\draw[-,thick]  (-0.5,1)--++(0.5,0);
	\draw[-,thick]  (-0.5,-0.7)--++(2,0);
	\draw[-,thick]  (-0.5,-1)--++(2,0);
	\draw[thick](-0.5,0) .. controls (-0.5-1/8,0) and (-0.5-2/8,-3/6) .. (-0.5,-0.7);
	\draw[thick](1.5,0) .. controls (1.5+1/8,0) and (1.5+2/8,-3/6) .. (1.5,-0.7);
}  
+
\diagram{
	\draw[-,thick]  (-1,0)--++(2,0);
	\draw[-,thick]  (-1,1)--++(2,0);
} \right],
\end{align}
where again we used $\text{Tr} \left[ \rho \right] = 1$. This expression amounts to a simple sum of all possible reduced density matrices. We can invert our diagram for the two qubit case by first tracing over one of its qubits to get
\begin{align}
	\diagram{
		\draw[fill=tensorblue, thick] (0,-0.5) rectangle (1,1.5) (C)  node[pos=0.5] {$S \left[ \rho \right] $};
		\draw[-,thick]  (1,0)--++(0.5,0);
		\draw[-,thick]  (1,1)--++(0.5,0);
		\draw[-,thick]  (-0.5,0)--++(0.5,0);
		\draw[-,thick]  (-0.5,1)--++(0.5,0);
		\draw[-,thick]  (-0.5,2)--++(2,0);
		\draw[thick](1.5,1) .. controls (1.5+1/4,1) and (1.5+2/4,3/4+1) .. (1.5,2);
		\draw[thick](-0.5,1) .. controls (-0.5-1/4,1) and (-0.5-2/4,3/4+1) .. (-0.5,2);
	}
=
\frac{1}{3} &\left[ 
\diagram{
	\draw[fill=tensorblue, thick] (0,-0.5) rectangle (1,1.5) (C)  node[pos=0.5] {$\rho$};
	\draw[-,thick]  (1,0)--++(0.5,0);
	\draw[-,thick]  (1,1)--++(0.5,0);
	\draw[-,thick]  (-0.5,0)--++(0.5,0);
	\draw[-,thick]  (-0.5,1)--++(0.5,0);
	\draw[-,thick]  (-0.5,2)--++(2,0);
	\draw[thick](1.5,1) .. controls (1.5+1/4,1) and (1.5+2/4,3/4+1) .. (1.5,2);
	\draw[thick](-0.5,1) .. controls (-0.5-1/4,1) and (-0.5-2/4,3/4+1) .. (-0.5,2);
}  \right.\nonumber\\
+
&\left.
\diagram{\draw[-,thick]  (0,0)--++(2,0);}
\right].
\end{align}
This allows us to rewrite the reduced density matrices in terms of the partially summed $S[\rho]$ . Using this relation, we then arrive at the inverse of the two-qubit classical shadow
\begin{align}
\diagram{
		\draw[fill=tensorblue, thick] (0,-0.5) rectangle (1,1.5) (C)  node[pos=0.5] {$\rho$};
		\draw[-,thick]  (1,0)--++(0.5,0);
		\draw[-,thick]  (1,1)--++(0.5,0);
		\draw[-,thick]  (-0.5,0)--++(0.5,0);
		\draw[-,thick]  (-0.5,1)--++(0.5,0);
	} =\: 
&9 \: \diagram{
	\draw[fill=tensorblue, thick] (0,-0.5) rectangle (1,1.5) (C)  node[pos=0.5] {$S \left[ \rho \right]$};
	\draw[-,thick]  (1,0)--++(0.5,0);
	\draw[-,thick]  (1,1)--++(0.5,0);
	\draw[-,thick]  (-0.5,0)--++(0.5,0);
	\draw[-,thick]  (-0.5,1)--++(0.5,0);
} - 
3\diagram{
	\draw[fill=tensorblue, thick] (0,-0.5) rectangle (1,1.5) (C)  node[pos=0.5] {$S \left[ \rho \right]$};
	\draw[-,thick]  (1,0)--++(0.5,0);
	\draw[-,thick]  (1,1)--++(0.5,0);
	\draw[-,thick]  (-0.5,0)--++(0.5,0);
	\draw[-,thick]  (-0.5,1)--++(0.5,0);
	\draw[-,thick]  (-0.5,1.7)--++(2,0);
	\draw[-,thick]  (-0.5,2)--++(2,0);
	\draw[thick] (-0.5,1) .. controls (-0.5-1/8,1) and (-0.5-2/8,3/6+1) .. (-0.5,1.7);
	\draw[thick] (1.5,1) .. controls (1.5+1/8,1) and (1.5+2/8,3/6+1) .. (1.5,1.7);
}  \nonumber \\
&-3\diagram{
	\draw[fill=tensorblue, thick] (0,-0.5) rectangle (1,1.5) (C)  node[pos=0.5] {$S \left[ \rho \right]$};
	\draw[-,thick]  (1,0)--++(0.5,0);
	\draw[-,thick]  (1,1)--++(0.5,0);
	\draw[-,thick]  (-0.5,0)--++(0.5,0);
	\draw[-,thick]  (-0.5,1)--++(0.5,0);
	\draw[-,thick]  (-0.5,-0.7)--++(2,0);
	\draw[-,thick]  (-0.5,-1)--++(2,0);
	\draw[thick] (-0.5,0) .. controls (-0.5-1/8,0) and (-0.5-2/8,-3/6) .. (-0.5,-0.7);
	\draw[thick] (1.5,0) .. controls (1.5+1/8,0) and (1.5+2/8,-3/6) .. (1.5,-0.7);
} 
+
 \diagram{
 	\draw[-,thick]  (-1,0)--++(2,0);
 	\draw[-,thick]  (-1,1)--++(2,0);
 }. 
\end{align}
As for the one qubit case, we can convert our diagrams back to algebraic expressions. The results of this two-qubit case amount to the forward expression 
\begin{equation}
S[\rho] = \frac{1}{9}(\rho + I\otimes \text{Tr}_1 \left[\rho\right] + \text{Tr}_2 \left[\rho\right]\otimes I + I\otimes I)
\end{equation}
and the inverse expression
\begin{equation}
    \rho = 9S\left[\rho\right] - 3I\otimes \text{Tr}_1 S\left[\rho\right] -3\text{Tr}_2 S\left[\rho\right]\otimes I + I\otimes I
\end{equation}
\emph{$N$-qubit case.} The formulas we have derived for the one- and two-qubit cases readily extend to $N$-qubits. They are
\begin{equation}
    S\left[\rho \right] = \frac{1}{3^L}\left(\rho + \sum_l  \text{Tr}_l \rho +
    \sum_{l\neq l'}\text{Tr}_{ll'}\rho + \ldots + I\right)
\end{equation}
for the forward map and for the inverse map
\begin{align}
    \label{eq:CSexpansion}
    &\rho = 3^L S\left[\rho \right]-
    3^{L-1}\sum_l \text{Tr}_l S\left[\rho \right] \nonumber \\
    &+ 3^{L-2}\sum_{l\neq l'}\text{Tr}_{ll'}S\left[\rho \right] +\ldots+(-1)^LI,
\end{align}
 
where we have suppressed the presence of identity operators that replace traced out regions for ease of notation. These expressions satisfy $\text{Tr}[\rho]=1$ and $S[\rho]\geq 0$. The inverse map is not non-negative in general, but should be for the density matrices resulting from this map.
Equation \ref{eq:CSexpansion} may also be written as:
\begin{equation}
    \rho = \frac{1}{N} \sum_{n=1}^N\bigotimes_{l=0}^{L} (3 \ket{b_l^{(n)};\sigma_l^{(n)}} \bra{b_l^{(n)};\sigma_l^{(n)}} -1 ),
\end{equation}
where $l$ denotes the site-index and $n$ denotes the shot-index.

\section{Symmetries of 1DTFIM}
\label{app:symmetry}
\subsection{$\mathbb{Z}_2$ symmetry}
The 1DTFIM is invariant under global flipping of the z-component of the spin, the $\mathbb{Z}_2$ symmetry. This unitary symmetry can be expressed as 
\begin{equation}
    \mathcal{S} =\prod_i X_i.
\end{equation}
\comment{$\mathcal{S} X_i \mathcal{S}^{-1} = X_i$, $\mathcal{S} Y_i \mathcal{S}^{-1} = Y_i$, and $\mathcal{S} Z_i \mathcal{S}^{-1} = -Z_i$, thus the action of $\mathcal{S}$ switches $\ket{0}$ and $\ket{1}$.} We can check that the symmetry operator $\mathcal{S}$ commutes with the TFIM Hamiltonian i.e. $\left[ \mathcal{S}, H_{\text{1DTFIM}}\right] = 0$. This allows us to write $H_{\text{1DTFIM}}$ in block diagonal form with each block corresponding to eigenvalues $1$ (even parity) and $-1$ (odd parity)of $\mathcal{S}$. States in these sectors evolve independently of each other. If we start in GHZ state
\begin{equation}
    \ket{\text{GHZ}} = \frac{\ket{00\cdots 0} + \ket{11 \cdots 1}}{\sqrt{2}},
\end{equation}
a time reversal even state, we remain in the even sector under time evolution. Since the magnetization of this state is $0$, the magnetization will stay at this value forever. Thus, equilibration of the order parameter is not an issue. Since the 1DTFIM Hamiltonian is purely real, it is symmetric under complex conjugation $\mathcal{K}$, and consequently the eigenvalues are also real. Hence, it is also symmetric under $\mathcal{T} = \mathcal{S} \mathcal{K}$ i.e. time reversal symmetry.

\subsection{Chiral symmetry}
The 1DTFIM is also symmetric under the following chiral operator
\begin{equation}
    \mathcal{C} = ZYZY\cdots ZY
\end{equation}
We can check that $C$ anticommutes with $H_{{\text{1DTFIM}}}$ i.e. $\{ \mathcal{C}, H_{\text{1DTFIM}}\} = 0$, so for every energy eigenstate $E$, there exists an eigenstate with $-E$, which makes the spectrum of 1DTFIM mirror symmetric about zero energy.

\section{Equilibration from  initial states}\label{app:equilibration}

\begin{figure}[ht]
\centering
\includegraphics[scale=1]{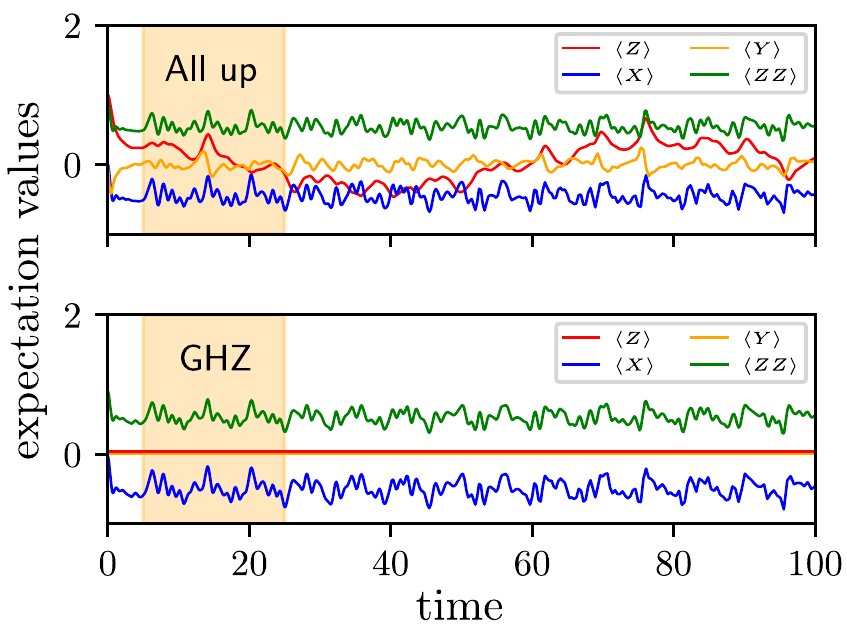}
\caption{Time evolution of $\expval{Z}$, $\expval{X}$ and $\expval{ZZ}$ for ferromagnetic (all up) and GHZ (all up plus all down) states for 10 sites and $h_x=0.8$. $t\in[5,25]$ (highlighted in orange) is the sampling window used for generating the TACS data. The order parameter $\expval{Z}$ for the ferromagnetic state doesn't equilibrate in this window}
\label{fig:equilibration}
\end{figure}

An important resource for studying microcanonical phases using quantum dynamics is an initial state that equilibrates within the time scale $T$ accessible to a quantum device. The initial state sets the energy and the symmetry sector of the microcanonical ensemble resulting from time-averaging over $[0,T]$. 

In figure \ref{fig:equilibration}, we present a numerical assessment of equilibration for some observables in the 1DTFIM. Specifically, we plot the evolution of expectation values for operators $\expval{Z}$,$\expval{X}$ and $\expval{ZZ}$ corresponding to two initial states: the ferromagnetic ($|00\ldots\rangle$) and GHZ ($\frac{1}{\sqrt{2}}(|00\ldots\rangle+|11\ldots\rangle)$) states respectively. We observe that the $\mathcal{T}$-odd operators such as $Z$ do not equilibrate for the all-up state within the sampling window whereas they are forced to be $0$ for the GHZ state by the $\mathcal{T}$-symmetry. Likewise, we find that $\mathcal{T}$-even operators such as $X$, $ZZ$ equilibrate and are identical for both initial states, also due to the $\mathcal{T}-symmetry$ (the ferromagnetic state is a superposition of $\mathcal{T}$-even and  $\mathcal{T}$-odd states, and the expectation value of $\mathcal{T}$-even operator for a $\mathcal{T}$-odd state yields 0).
Finally, we observe all equilibrating observables equilibrate within time scale of 5. For this reason, all of the dynamics results presented in the main manuscript used this numerical evidence time interval to $[5,25]$ (shown in the highlighted region in Fig. \ref{fig:equilibration} for time averages. Hence, we find numerical evidence for equilibration of local observables, evidence that formed an important basis upon which we carried out our dynamics simulations.

\section{Interpreting diffusion maps }   
\subsection{Learning physical parameters from 1DTFIM Ground State Data} 
The UL model we implemented was able to unveil the symmetry-breaking phase transition of 1DTFIM from ground state CS data (Fig. \ref{ground}(d)). It did so by generating diffusion coordinates that are related to relevant parameters, the order parameter $M_{z}$, and the model parameter $h_{x}$ of 1DTFIM. Fig. \ref{fig:correlation} shows the correlation between the diffusion coordinates and these parameters. 

% ground state parameter learning results begin
\begin{figure}[ht]
\centering
\includegraphics[width=0.46\textwidth,height=0.13\textheight]{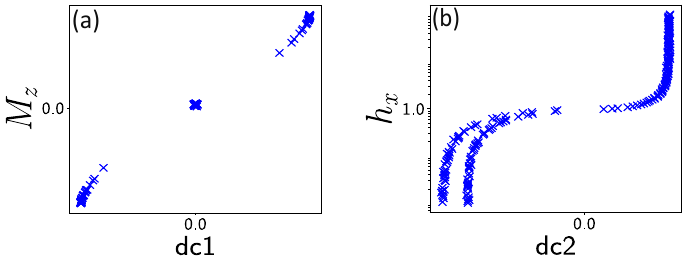}
\caption{Learning 1DTFIM model parameter and order parameter from ground state CS data. Correlation between (a) dc1 and order parameter $M_{z}$ and (b) dc2 and model parameter $h_{x}$.}
\label{fig:correlation}    
\end{figure}
% end

\subsection{Quantum Criticality in TACS Kernel Matrix}
\label{appendix_crit}

% quantum critical phase demonstration plot
\begin{figure}[ht]
\centering
\includegraphics[width=0.46\textwidth,height=0.3\textheight]{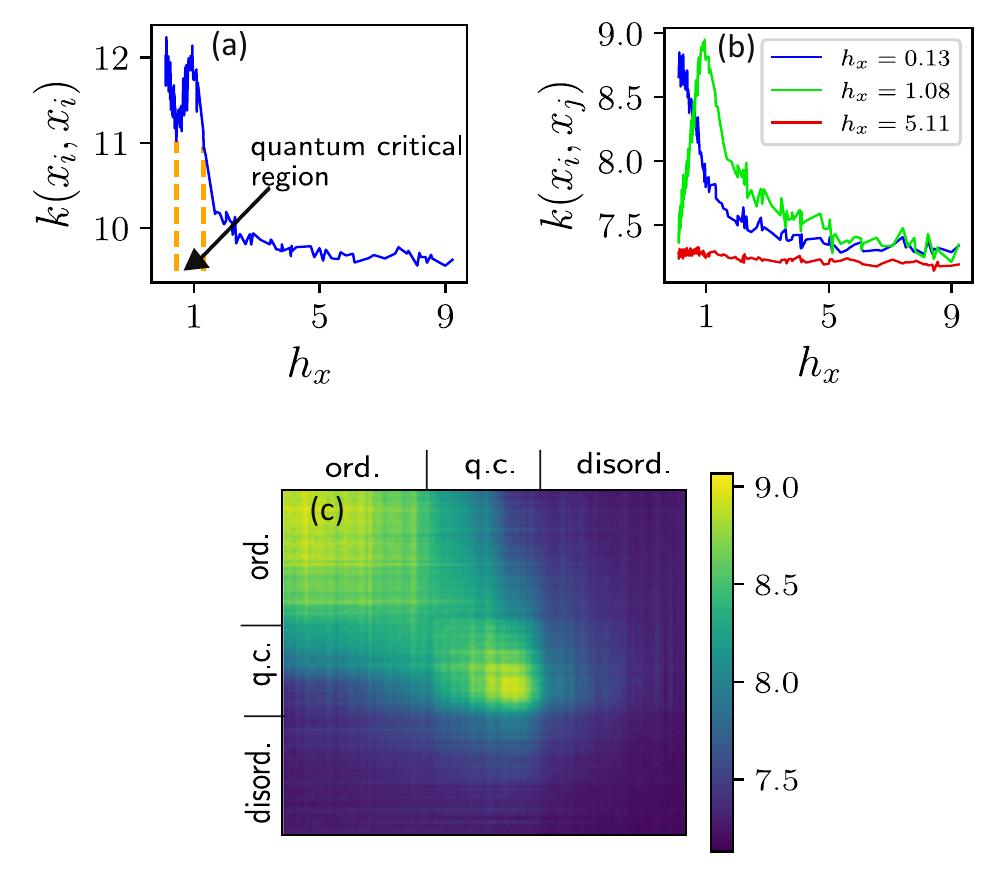}
\caption{The three distinct regions of the microcanonical phase diagram depicted via the kernel matrix $K$. (a) shows the diagonal elements of $K$ against their respective $h_{x}$ values. The quantum critical region is inferred from this plot. (b) shows three different rows of $K$, each belonging to a different region in the microcanonical phase diagram. (c) portrays the full matrix $K$ itself, where the distinct regions (ordered, quantum critical, and disordered) can be seen. (diagonal elements have been ignored in (b) and (c) for better visualization.) }
\label{fig:quantumcritical}   
\end{figure}
% quantum critical plot end

The microcanonical phase diagram being studied here has three characteristically distinct regions\textemdash namely, the ordered phase, the quantum critical region, and the disordered phase. The quantum critical region, although does not include the phase transition point, exhibits critical behavior characterized by singularity in the order parameter and the response functions. This behavior manifests itself in the feature space accompanying the shadows kernel function [Eq.~\ref{eq2}] as it contains the polynomial expansion of the reduced density matrices~\cite{huang2021provably} and can be analyzed from the kernel matrix.

Figure ~\ref{fig:quantumcritical} displays how the microcanonical phases reveal themselves in the kernel matrix. Figure~\ref{fig:quantumcritical}(a) shows a maximum in the diagonal elements of the kernel matrix in the quantum critical region due to increased correlation length. In the figure, we delineate the quantum critical region in the neighborhood of this peak. Likewise, Fig. ~\ref{fig:quantumcritical}(b) and (c) demonstrate how the kernel function between states behaves in different regions. The ordered states have low entropy; hence, greater "similarity" among themselves makes the kernel function take a higher value than the other regions, dropping sharply as we go out of that region. The disordered states have roughly uniform values for the kernel function with all other states due to their high entropy, and a peak in the critical region as discussed above. These character traits of each of these regions help us identify them from the kernel matrix. However, we don't use the kernel matrix for phase classification. We let the probabilities diffuse and use the diffusion matrix and resulting diffusion coordinates.     

\subsection{Qualitative similarity between dc's and susceptibility}
\label{sec:susceptibilities}
Susceptibility is an important quantity of interest to us because it diverges at the critical point. Although our microcanonical dynamics takes place at an energy above the ground state, we expect the signature of this divergence to be present in the quantum critical region. With the experience that diffusion coordinates correspond to phase-defining observables in the case of ground states, we plotted the the xx, yy and zz components of the susceptibility in Fig. \ref{fig:susceptibilities} 
computed using the 100k shot TACS dataset for the 1DTFIM to compare against dc1 and dc2. Although $\chi_{zz}$, a natural candidate for dc1, looks qualitatively similar to dc1 when inverted, we find that the $2^{nd}$ Renyi entropy is a better fit. Similarly, $\chi_{xx}$ behaves qualitatively similar to dc2 in sense that both are sharply peaked at the critical value of $h_x=1$. However, they do not share the same critical exponents, hence, we cannot make as strong of a claim as for dc1.

\begin{figure}[ht]
\centering
\includegraphics[scale=1]{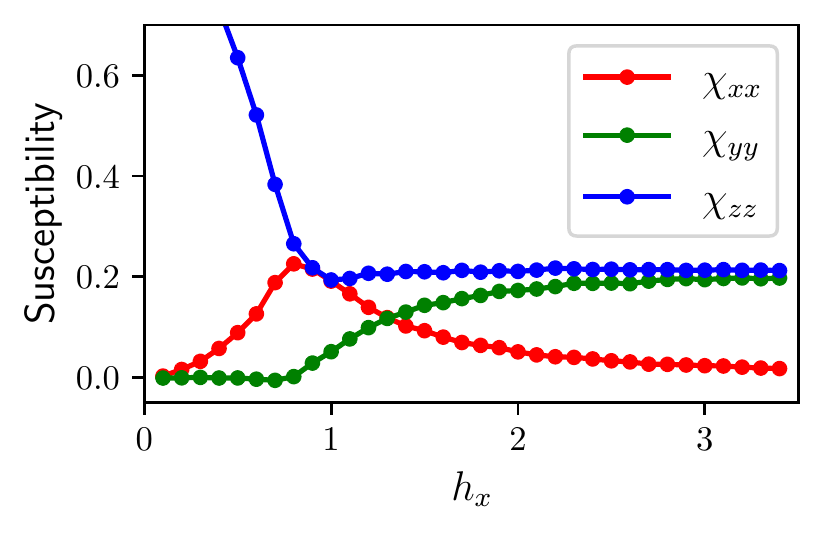}
\caption{xx, yy and zz components of the magnetic susceptibility for a 10-site 1DTFIM, computed using 10000 shot TACS dataset.}
\label{fig:susceptibilities}
\end{figure}

\subsection{Estimating The Critical Exponent from The TACS Diffusion Coordinates}
\label{app:nu}
In Fig. \ref{fig:nu_C}(a) below, we see that the second diffusion coordinate \text{dc2} in TACS diffusion maps approximates a power law in the quantum critical region. In order to estimate the critical exponent, we modeled \text{dc2} as:
\begin{equation}
\text{dc2}(h_x;a,p) = a\left|h_x-1\right|^{-p} + C
\end{equation}
where $a$ and $p$ are fitting parameters and $\nu$ is the critical exponent. 

It is evident that our estimate of $p$ depends on our choice of $C$. Fig. \ref{fig:nu_C} shows that dependence on either side of the critical point ($h_x=1$). We can obtain a probability distribution $P(C)$ on $C$ by modeling the Bayesian estimate of the 2nd Renyi entropy $S_2/n$ as a function of $dc1$:
\begin{equation}
S_2 / n(h_x;\alpha,C)  = \alpha (dc1\big(h_x\big) - C)
\end{equation}
Here, $\alpha$ and $C$ are fitting parameters. The ordinary least squares fit gives us the optimum value for $C$ ($C_{opt}$) with the least square error($\epsilon$). We then model $P(C)$ as a normal distribution, $P(C) = \mathcal{N}(C_{opt},\sigma=\epsilon^2)$ and plot it together with the dependence of $p$ on the shift $C$ in Fig. \ref{fig:nu_C} to visualize how an error in $C$ translates to an error in $\nu$.

% begin: critical exponent figure
\begin{figure}[ht]
\centering
\includegraphics[width=0.48\textwidth,height=0.6\textheight]{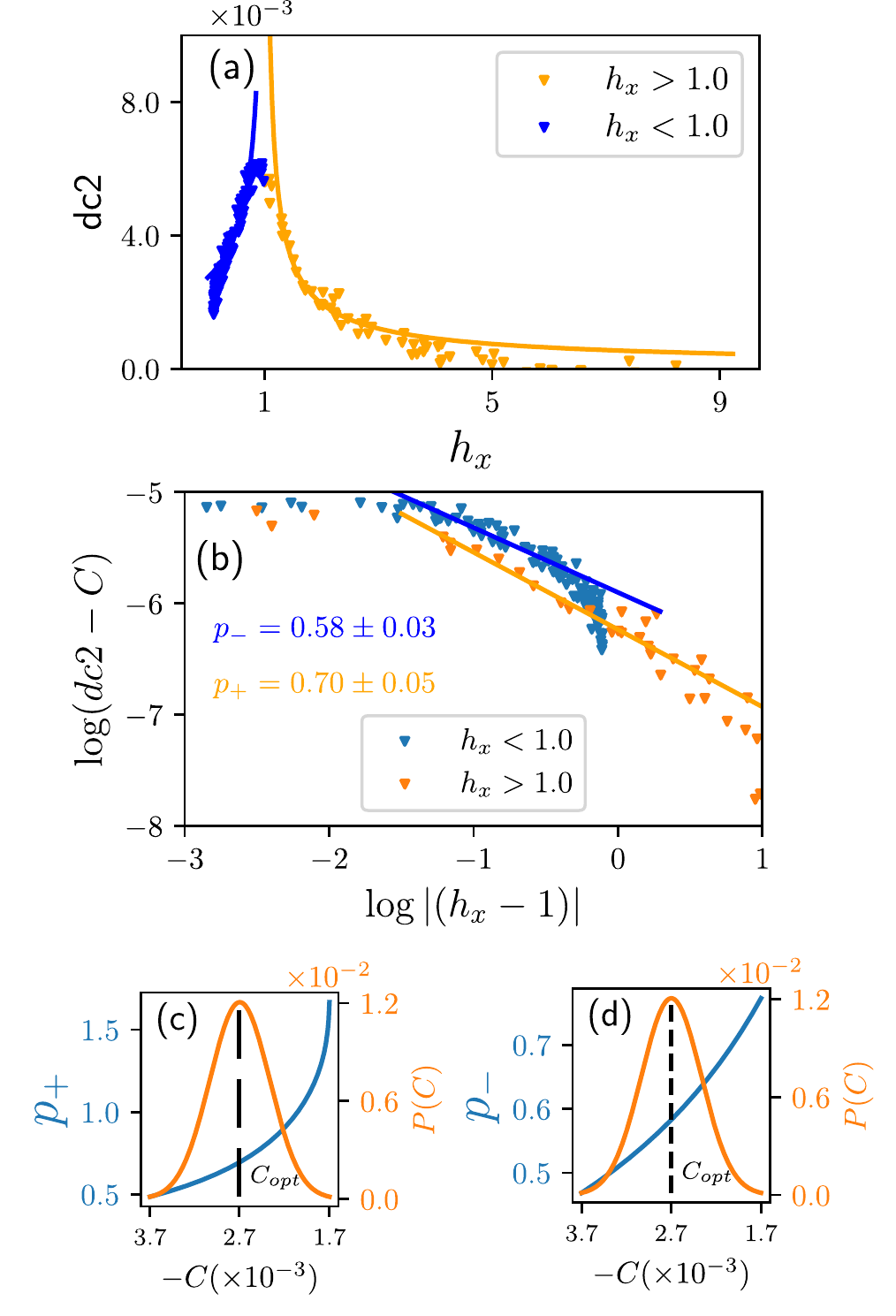}
\caption{Modeling the power-law behavior of dc2 in the critical region. (a) shows the dc2 value of each TACS plotted against their respective $h_x$ values, as well as the power law fit on either side of the critical point. (b) log-log plot of $|h_x-1|$ vs $(\text{dc2} - C)$, along with straight-line fits, the slopes give us the $p$ values .Here $C=-0.0027$. (c) and (d) show the dependence of the (c) $p_+$ and (d) $p_-$, estimated power law exponents on the shift in the diffusion coordinates $C$ discussed in section \ref{sec:ML}(blue) [dashed line shows the optimum $C$-value, which was chosen in (a) and (b)], along with the modeled normal distribution on $C$ obtained from error estimates on the shift needed to render dc1 positive near $h_x = 0$ (orange). Chosen range of $C$ is [$C_{opt}-3\sigma,C_{opt}+3\sigma$]}.
\label{fig:nu_C}
\end{figure}
% end: critical exponent figure

\section{Bayesian Inference Extrapolation of Entropy Data}\label{app:bi}
\begin{figure*}[ht]
\begin{center}
\includegraphics[scale=0.95]{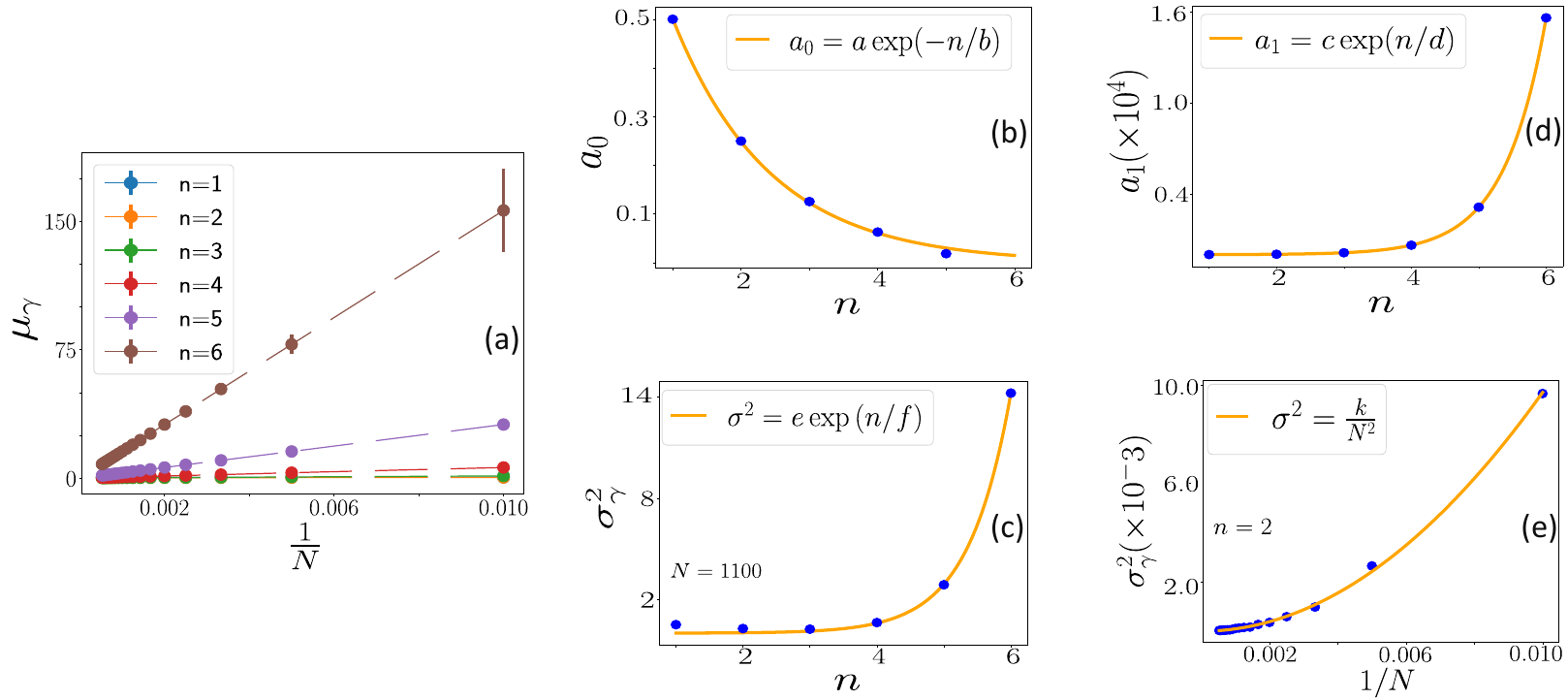}
\caption{Developing a model for purity $\gamma = \mathcal{N}(\mu_\gamma,\sigma_\gamma)$ from CS data on the maximally mixed state. (a) $\mu_\gamma$ is linear in $\frac{1}{N}$, so it takes the form: $\mu_{\gamma}=a_{0}(n) + \frac{a_{1}(n)}{N}$. (b) The intercept $a_{0}(n)$ can be modeled as an inverse exponential function of $n$, $a_0(n) = ae^{-bn}$ (c) The slope, $a_{1}(n)$, can be modeled as an exponential in $n$, $a_1(n) = ce^{dn}$. (d), (e) show that the variance $\sigma_\gamma^{2}$ can be closely approximated by an exponential function in $n$ and linear in $\frac{1}{N^{2}}$, $\sigma_\gamma^2 = ee^{fn}/N^2$.}
\label{Fig:bayes_model}
\end{center}
\end{figure*}

To infer the entropy in the limit of large $N$, the number of shots, and large $n$, the number of qubits, we need to extrapolate the estimates we obtain from CS data. At first glance, this would seem hard to do because the error in our estimates grows exponentially with the locality of the observables and entropy is not a local observable. However, the dynamics data we have obtained represents a mixed state with volume-law entanglement and, due to the finite energy---the microcanonical stand-in for temperature---typically has exponentially decaying correlations beyond a correlation length. Hence, we expect the entropy of the reduced density matrix of a region $A$ with $n_A$ qubits will obey $S \propto n_A$ even for small $n_A$. Our approach to extrapolate the entropy is therefore to build a probability model $p(X|\theta)$ with parameters $\theta$ that captures our estimated entropy value data $X$.

To simplify the calculation of entropy, we will compute the second Renyi entropy of a sub-region A: $S_A = -\log_2 \gamma_A$, $\gamma = Tr \rho_A^2$. To obtain a model of this entropy as a function of the number of CS shots $N$ and qubits $n_A$, a data set of values $Y$ given by the entropy $S_A$ and dependent variables $X$ given by $(N,n_A)$, we generated data from the maximally mixed state $\rho = (1/d)I$, $d=2^n_A$ the size of the Hilbert space. The results fit a model of the form (see Fig. \ref{Fig:bayes_model}):
\begin{equation}
    \mu_{\gamma}(n,N) = ae^{-bn} + ce^{dn}/N,\quad
    \sigma_\gamma = ee^{fn}
\end{equation}
with positive parameters $a$, $b$, $c$, $d$, $e$, $f$. Namely, we found the purity $\gamma$ is linear in $1/N$ but exponential in $n$. 
%\michael{Mabrur: continue this discussion, presenting your results including figures for the maximally mixed state showing that the mean values $\mu(X;\theta)$ are linear in $1/N$ with slope and intercept that are exponential in $n_A$, and variance $\sigma^2(X;\theta)$ that is proportional to $1/N$ with coefficient exponential in $n_A$}. 

Given the mean and variance as modeled above, we can then model the probability distribution from which a given data point $(\vec x,y)\in (X,Y)$ is a sample as a Gaussian:
\begin{equation}
    P(y\backslash\theta,\vec x) = \frac{1}{\sqrt{2\pi\sigma^2(\vec x;\theta)}}
    e^{-\left(y-\mu(\vec x;\theta)\right)^2/2\sigma^2(\vec x;\theta)}.
\end{equation}
Then by Bayes Law, we can learn the posterior
\begin{equation}
    p(\theta\backslash Y,X) = \frac{\prod_{(\vec x,y)\in (X,Y)} P(y\backslash\theta,\vec x)P(\theta)}{P(Y\backslash X)}
\end{equation}
where $P(X,Y) = \int d\theta \prod_{(X,Y)} P(\vec x,y\backslash\theta)P(\theta)$ is called the evidence that provides a sense of how well the model is performing. 

The probability of observing a new data point $(\vec x',y')$ is then given by the posterior predictive
\begin{equation}
    P(\vec x',y\backslash X,Y) = \int d\theta p(\vec x',y'\backslash\theta)P(\theta\backslash X,Y)
\end{equation}
An estimate of which is obtainable from a set of samples $\Theta$ drawn from $P(\theta\backslash X,Y)$
\begin{equation}\label{eq:posteriorpredict}
    P(\vec x',y\backslash X,Y) = \frac{1}{|\Theta|} \sum_{\theta\in\Theta} p(\vec x',y'\backslash\theta)
\end{equation}
We are then specifically interested in the mean  and standard deviation of $P((\infty,\infty),y\backslash X,Y)$. Knowing this, we solve the problem of extrapolating the entropy from a finite number of shots and qubits for the entropy is the mean and our uncertainty in obtaining it is the standard deviation.

It remains then to obtain samples from the posterior $P(\theta\backslash X, Y)$. We could do so using a straightforward Montecarlo algorithm. For example, starting with an initial choice for the parameters $\theta_0$, we pick a random direction in parameter space and move an amount $\delta$ in that direction to obtain $\theta_{trial}$. We then compute
\begin{multline}
    \log(r) = \log \frac{P(\theta_{trial}\backslash X,Y)}{P(\theta_0\backslash X,Y)} = \\\sum_{(\vec x,y)\in(X,Y)}\left(\log P(\vec x,y\backslash\theta_{trial}) - \log P(\vec x,y\backslash\theta_0)\right) +\\ \log P(\theta_{trial}) - \log P(\theta_0)
\end{multline}
which simplifies if we choose a uniform distribution for $P(\theta)$. We keep the trial, setting $\theta_1 =\theta_{trial}$ if a random number $q$ between 0 and 1 satisfies $q<r$ and reject otherwise. Either way, we repeat the process generating ultimately a list $\Theta$ of correlated samples $\theta_i$ from which we can estimate the entropy and uncertainty from $P((\infty,\infty),y\backslash X, Y)$.

However, a better approach than the Metropolis algorithm is to use the NUTS algorithm available in PyMC instead. This algorithm automatically chooses parameters in hamiltonian monte-carlo (HMC) and is more efficient than Metropolis for Bayesian inference. See Ref. \onlinecite{salvatier2016probabilistic}.

\bibliographystyle{unsrt}
\bibliography{main.bib}

\begin{thebibliography}{10}

\bibitem{Cao_2019}
Yudong Cao, Jonathan Romero, Jonathan~P. Olson, Matthias Degroote, Peter~D.
  Johnson, M{\'a}ria Kieferov{\'a}, Ian~D. Kivlichan, Tim Menke, Borja
  Peropadre, Nicolas P.~D. Sawaya, Sukin Sim, Libor Veis, and Al{\'a}n
  Aspuru-Guzik.
\newblock Quantum chemistry in the age of quantum computing.
\newblock {\em Chemical Reviews}, 119(19):10856--10915, 10 2019.

\bibitem{Wecker_2015}
Dave Wecker, Matthew~B. Hastings, Nathan Wiebe, Bryan~K. Clark, Chetan Nayak,
  and Matthias Troyer.
\newblock Solving strongly correlated electron models on a quantum computer.
\newblock {\em Phys. Rev. A}, 92:062318, Dec 2015.

\bibitem{Ma_2020}
He~Ma, Marco Govoni, and Giulia Galli.
\newblock Quantum simulations of materials on near-term quantum computers.
\newblock {\em npj Computational Materials}, 6(1):85, 2020.

\bibitem{highTc_1}
Nagaosa Naoto.
\newblock Superconductivity and antiferromagnetism in high-<i>t</i><sub>c</sub>
  cuprates.
\newblock {\em Science}, 275(5303):1078--1079, 1997.

\bibitem{highTc_2}
Xingjiang Zhou, Wei-Sheng Lee, Masatoshi Imada, Nandini Trivedi, Philip
  Phillips, Hae-Young Kee, P{\"a}ivi T{\"o}rm{\"a}, and Mikhail Eremets.
\newblock High-temperature superconductivity.
\newblock {\em Nature Reviews Physics}, 3(7):462--465, 2021.

\bibitem{mcgreevy_2010}
John McGreevy.
\newblock In pursuit of a nameless metal, Oct 2010.

\bibitem{Laughlin_1999}
R.~B. Laughlin.
\newblock Nobel lecture: Fractional quantization.
\newblock {\em Rev. Mod. Phys.}, 71:863--874, Jul 1999.

\bibitem{Savary_2016}
Lucile Savary and Leon Balents.
\newblock Quantum spin liquids: a review.
\newblock {\em Reports on Progress in Physics}, 80(1):016502, nov 2016.

\bibitem{dmrg1}
Steven~R. White.
\newblock Density matrix formulation for quantum renormalization groups.
\newblock {\em Phys. Rev. Lett.}, 69:2863--2866, Nov 1992.

\bibitem{dmrg2}
Steven~R. White.
\newblock Density-matrix algorithms for quantum renormalization groups.
\newblock {\em Phys. Rev. B}, 48:10345--10356, Oct 1993.

\bibitem{dmrg3}
Ulrich Schollwöck.
\newblock The density-matrix renormalization group in the age of matrix product
  states.
\newblock {\em Annals of Physics}, 326(1):96--192, 2011.
\newblock January 2011 Special Issue.

\bibitem{qmc}
Federico Becca and Sandro Sorella.
\newblock {\em Quantum Monte Carlo Approaches for Correlated Systems}.
\newblock Cambridge University Press, 2017.

\bibitem{dmft1}
Antoine Georges, Gabriel Kotliar, Werner Krauth, and Marcelo~J. Rozenberg.
\newblock Dynamical mean-field theory of strongly correlated fermion systems
  and the limit of infinite dimensions.
\newblock {\em Rev. Mod. Phys.}, 68:13--125, Jan 1996.

\bibitem{Feynman1982}
Richard~P. Feynman.
\newblock Simulating physics with computers.
\newblock {\em International Journal of Theoretical Physics}, 21(6):467--488,
  1982.

\bibitem{kitaev2002classical}
Alexei~Yu Kitaev, Alexander Shen, and Mikhail~N Vyalyi.
\newblock {\em Classical and quantum computation}.
\newblock Number~47. American Mathematical Soc., 2002.

\bibitem{Gorman_2021}
Bryan O'Gorman, Sandy Irani, James Whitfield, and Bill Fefferman.
\newblock Electronic structure in a fixed basis is qma-complete, 2021.

\bibitem{Gharibian2019complexity}
Sevag Gharibian and Justin Yirka.
\newblock The complexity of simulating local measurements on quantum systems.
\newblock {\em {Quantum}}, 3:189, September 2019.

\bibitem{Bookatz2014QMA}
Adam~D. Bookatz.
\newblock Qma-complete problems.
\newblock {\em Quantum Info. Comput.}, 14:361–383, apr 2014.

\bibitem{kemp2004complexity}
Julia Kempe, Alexei Kitaev, and Oded Regev.
\newblock The complexity of the local hamiltonian problem.
\newblock In Kamal Lodaya and Meena Mahajan, editors, {\em FSTTCS 2004:
  Foundations of Software Technology and Theoretical Computer Science}, pages
  372--383, Berlin, Heidelberg, 2005. Springer Berlin Heidelberg.

\bibitem{Baez2020dynamical}
Maria~Laura Baez, Marcel Goihl, Jonas Haferkamp, Juani Bermejo-Vega, Marek
  Gluza, and Jens Eisert.
\newblock Dynamical structure factors of dynamical quantum simulators.
\newblock {\em Proceedings of the National Academy of Sciences},
  117(42):26123–26134, Oct 2020.

\bibitem{Rudi2020approximating}
Alessandro Rudi, Leonard Wossnig, Carlo Ciliberto, Andrea Rocchetto,
  Massimiliano Pontil, and Simone Severini.
\newblock Approximating hamiltonian dynamics with the nyström method.
\newblock {\em Quantum}, 4:234, Feb 2020.

\bibitem{Aaronson2018}
Scott Aaronson.
\newblock Shadow tomography of quantum states.
\newblock In {\em Proceedings of the 50th Annual ACM SIGACT Symposium on Theory
  of Computing}, STOC 2018, page 325–338, New York, NY, USA, 2018.
  Association for Computing Machinery.

\bibitem{huang2020predicting}
Hsin-Yuan Huang, Richard Kueng, and John Preskill.
\newblock Predicting many properties of a quantum system from very few
  measurements.
\newblock {\em Nature Physics}, 16(10):1050–1057, Jun 2020.

\bibitem{huang2021provably}
Hsin-Yuan Huang, Richard Kueng, Giacomo Torlai, Victor~V. Albert, and John
  Preskill.
\newblock Provably efficient machine learning for quantum many-body problems.
\newblock {\em Science}, 377(6613):eabk3333, 2022.

\bibitem{Huang2022Learning}
Hsin-Yuan Huang, Sitan Chen, and John Preskill.
\newblock Learning to predict arbitrary quantum processes, 2022.

\bibitem{von2010proof}
John von Neumann.
\newblock Proof of the ergodic theorem and the h-theorem in quantum mechanics.
\newblock {\em The European Physical Journal H}, 35(2):201--237, 2010.

\bibitem{Ahranov_2013}
Dorit Aharonov, Itai Arad, and Thomas Vidick.
\newblock The quantum pcp conjecture.
\newblock 2013.

\bibitem{AnshuNLTS}
Anurag Anshu, Nikolas~P. Breuckmann, and Chinmay Nirkhe.
\newblock Nlts hamiltonians from good quantum codes, 2022.

\bibitem{Wilming_towards_rapid_eq_2017}
H.~Wilming, M.~Goihl, C.~Krumnow, and J.~Eisert.
\newblock Towards local equilibration in closed interacting quantum many-body
  systems, 2017.

\bibitem{Malabara_rapid_eq}
Artur S.~L. Malabarba, Luis~Pedro Garc\'{\i}a-Pintos, Noah Linden, Terence~C.
  Farrelly, and Anthony~J. Short.
\newblock Quantum systems equilibrate rapidly for most observables.
\newblock {\em Phys. Rev. E}, 90:012121, Jul 2014.

\bibitem{Hetterich2015}
Daniel Hetterich, Moritz Fuchs, and Bj\"orn Trauzettel.
\newblock Equilibration in closed quantum systems: Application to spin qubits.
\newblock {\em Phys. Rev. B}, 92:155314, Oct 2015.

\bibitem{coifman2006diffusion}
Ronald~R Coifman and St{\'e}phane Lafon.
\newblock Diffusion maps.
\newblock {\em Applied and computational harmonic analysis}, 21(1):5--30, 2006.

\bibitem{de2008introduction}
J~De~la Porte, BM~Herbst, W~Hereman, and SJ~Van Der~Walt.
\newblock An introduction to diffusion maps.
\newblock In {\em Proceedings of the 19th symposium of the pattern recognition
  association of South Africa (PRASA 2008), Cape Town, South Africa}, pages
  15--25, 2008.

\bibitem{randomized_toolbox_2022}
Andreas Elben, Steven~T. Flammia, Hsin-Yuan Huang, Richard Kueng, John
  Preskill, Benoît Vermersch, and Peter Zoller.
\newblock The randomized measurement toolbox, 2022.

\bibitem{Knill_2005}
E.~Knill.
\newblock Quantum computing with realistically noisy devices.
\newblock {\em Nature}, 434(7029):39--44, mar 2005.

\bibitem{nielsen_chuang_2010}
Michael~A. Nielsen and Isaac~L. Chuang.
\newblock {\em Quantum Computation and Quantum Information: 10th Anniversary
  Edition}.
\newblock Cambridge University Press, 2010.

\bibitem{goldstein2010normal}
Sheldon Goldstein, Joel~L Lebowitz, Christian Mastrodonato, Roderich Tumulka,
  and Nino Zangh{\`\i}.
\newblock Normal typicality and von neumann’s quantum ergodic theorem.
\newblock {\em Proceedings of the Royal Society A: Mathematical, Physical and
  Engineering Sciences}, 466(2123):3203--3224, 2010.

\bibitem{Gogolin_2016}
Christian Gogolin and Jens Eisert.
\newblock Equilibration, thermalisation, and the emergence of statistical
  mechanics in closed quantum systems.
\newblock {\em Reports on Progress in Physics}, 79(5):056001, apr 2016.

\bibitem{Short_2012}
Anthony~J Short and Terence~C Farrelly.
\newblock Quantum equilibration in finite time.
\newblock {\em New Journal of Physics}, 14(1):013063, jan 2012.

\bibitem{yunger2016microcanonical}
Nicole Yunger~Halpern, Philippe Faist, Jonathan Oppenheim, and Andreas Winter.
\newblock Microcanonical and resource-theoretic derivations of the thermal
  state of a quantum system with noncommuting charges.
\newblock {\em Nature communications}, 7(1):1--7, 2016.

\bibitem{vinjanampathy2016quantum}
Sai Vinjanampathy and Janet Anders.
\newblock Quantum thermodynamics.
\newblock {\em Contemporary Physics}, 57(4):545--579, 2016.

\bibitem{deutsch1991quantum}
Josh~M Deutsch.
\newblock Quantum statistical mechanics in a closed system.
\newblock {\em Physical review a}, 43(4):2046, 1991.

\bibitem{srednicki1994chaos}
Mark Srednicki.
\newblock Chaos and quantum thermalization.
\newblock {\em Physical review e}, 50(2):888, 1994.

\bibitem{gong2022bounds}
Zongping Gong and Ryusuke Hamazaki.
\newblock Bounds in nonequilibrium quantum dynamics.
\newblock {\em arXiv preprint arXiv:2202.02011}, 2022.

\bibitem{margolus2021counting}
Norman Margolus.
\newblock Counting distinct states in physical dynamics.
\newblock {\em arXiv preprint arXiv:2111.00297}, 2021.

\bibitem{itensor}
Matthew Fishman, Steven~R. White, and E.~Miles Stoudenmire.
\newblock The \mbox{ITensor} software library for tensor network calculations,
  2020.

\bibitem{Huang2019predicting}
Hsin-Yuan Huang and Richard Kueng.
\newblock Predicting features of quantum systems from very few measurements,
  2019.

\bibitem{PreskillNISQ}
John Preskill.
\newblock Quantum {C}omputing in the {NISQ} era and beyond.
\newblock {\em {Quantum}}, 2:79, August 2018.

\bibitem{carleo2019machine}
Giuseppe Carleo, Ignacio Cirac, Kyle Cranmer, Laurent Daudet, Maria Schuld,
  Naftali Tishby, Leslie Vogt-Maranto, and Lenka Zdeborov{\'a}.
\newblock Machine learning and the physical sciences.
\newblock {\em Reviews of Modern Physics}, 91(4):045002, 2019.

\bibitem{rodriguez2019identifying}
Joaquin~F Rodriguez-Nieva and Mathias~S Scheurer.
\newblock Identifying topological order through unsupervised machine learning.
\newblock {\em Nature Physics}, 15(8):790--795, 2019.

\bibitem{lidiak2020unsupervised}
Alexander Lidiak and Zhexuan Gong.
\newblock Unsupervised machine learning of quantum phase transitions using
  diffusion maps.
\newblock {\em Physical Review Letters}, 125(22):225701, 2020.

\bibitem{borg2005modern}
Ingwer Borg and Patrick~JF Groenen.
\newblock {\em Modern multidimensional scaling: Theory and applications}.
\newblock Springer Science \& Business Media, 2005.

\bibitem{sachdev_2011}
Subir Sachdev.
\newblock {\em Quantum Phase Transitions}.
\newblock Cambridge University Press, 2 edition, 2011.

\bibitem{wu2018crossovers}
Jianda Wu, Lijun Zhu, and Qimiao Si.
\newblock Crossovers and critical scaling in the one-dimensional
  transverse-field ising model.
\newblock {\em Physical Review B}, 97(24):245127, 2018.

\bibitem{tdvp}
Jutho Haegeman, Christian Lubich, Ivan Oseledets, Bart Vandereycken, and Frank
  Verstraete.
\newblock Unifying time evolution and optimization with matrix product states.
\newblock {\em Phys. Rev. B}, 94:165116, Oct 2016.

\bibitem{timeevomps}
https://github.com/orialb/TimeEvoMPS.jl.

\bibitem{TACSgithub}
\url{https://github.com/Lawler-Research-Group/TACS}.

\bibitem{cardy1996scaling}
John Cardy.
\newblock {\em Scaling and renormalization in statistical physics}, volume~5.
\newblock Cambridge university press, 1996.

\bibitem{wiki:ising}
{Wikipedia contributors}.
\newblock Ising critical exponents, 2022.
\newblock [Online; accessed 12-Dec-2022].

\bibitem{Wu_2011_entropyaccumulation}
Jianda Wu, Lijun Zhu, and Qimiao Si.
\newblock Entropy accumulation near quantum critical points: effects beyond
  hyperscaling.
\newblock {\em Journal of Physics: Conference Series}, 273(1):012019, jan 2011.

\bibitem{lukens2021bayesian}
Joseph~M Lukens, Kody~JH Law, and Ryan~S Bennink.
\newblock A bayesian analysis of classical shadows.
\newblock {\em npj Quantum Information}, 7(1):1--10, 2021.

\bibitem{Shack}
Stefan~H. Sack, Raimel~A. Medina, Alexios~A. Michailidis, Richard Kueng, and
  Maksym Serbyn.
\newblock Avoiding barren plateaus using classical shadows.
\newblock {\em PRX Quantum}, 3:020365, Jun 2022.

\bibitem{Yao_2021}
Yong-Xin Yao, Niladri Gomes, Feng Zhang, Cai-Zhuang Wang, Kai-Ming Ho, Thomas
  Iadecola, and Peter~P. Orth.
\newblock Adaptive variational quantum dynamics simulations.
\newblock {\em PRX Quantum}, 2:030307, Jul 2021.

\bibitem{Yuan_2019}
Xiao Yuan, Suguru Endo, Qi~Zhao, Ying Li, and Simon~C. Benjamin.
\newblock Theory of variational quantum simulation.
\newblock {\em Quantum}, 3:191, oct 2019.

\bibitem{VFF_2020}
Cristina C{\^\i}rstoiu, Zo{\"e} Holmes, Joseph Iosue, Lukasz Cincio, Patrick~J.
  Coles, and Andrew Sornborger.
\newblock Variational fast forwarding for quantum simulation beyond the
  coherence time.
\newblock {\em npj Quantum Information}, 6(1):82, 2020.

\bibitem{Predicting_Gibbs}
Luuk Coopmans, Yuta Kikuchi, and Marcello Benedetti.
\newblock Predicting gibbs state expectation values with pure thermal shadows,
  2022.

\bibitem{QITE}
Mario Motta, Chong Sun, Adrian T.~K. Tan, Matthew~J. O'Rourke, Erika Ye,
  Austin~J. Minnich, Fernando G. S.~L. Brand{\~a}o, and Garnet Kin-Lic Chan.
\newblock Determining eigenstates and thermal states on a quantum computer
  using quantum imaginary time evolution.
\newblock {\em Nature Physics}, 16(2):205--210, 2020.

\bibitem{Partition_zeros}
Akhil Francis, Daiwei Zhu, Cinthia~Huerta Alderete, Sonika Johri, Xiao Xiao,
  James~K. Freericks, Christopher Monroe, Norbert~M. Linke, and Alexander~F.
  Kemper.
\newblock Many-body thermodynamics on quantum computers via partition function
  zeros.
\newblock {\em Science Advances}, 7(34):eabf2447, 2021.

\bibitem{Rydberg_atoms}
Hannes Bernien, Sylvain Schwartz, Alexander Keesling, Harry Levine, Ahmed
  Omran, Hannes Pichler, Soonwon Choi, Alexander~S. Zibrov, Manuel Endres,
  Markus Greiner, Vladan Vuleti{\'c}, and Mikhail~D. Lukin.
\newblock Probing many-body dynamics on a 51-atom quantum simulator.
\newblock {\em Nature}, 551(7682):579--584, 2017.

\bibitem{Karamlou2022}
Amir~H. Karamlou, Jochen Braum{\"u}ller, Yariv Yanay, Agustin Di~Paolo,
  Patrick~M. Harrington, Bharath Kannan, David Kim, Morten Kjaergaard,
  Alexander Melville, Sarah Muschinske, Bethany~M. Niedzielski, Antti
  Veps{\"a}l{\"a}inen, Roni Winik, Jonilyn~L. Yoder, Mollie Schwartz, Charles
  Tahan, Terry~P. Orlando, Simon Gustavsson, and William~D. Oliver.
\newblock Quantum transport and localization in 1d and 2d tight-binding
  lattices.
\newblock {\em npj Quantum Information}, 8(1):35, 2022.

\bibitem{salvatier2016probabilistic}
John Salvatier, Thomas~V Wiecki, and Christopher Fonnesbeck.
\newblock Probabilistic programming in python using pymc3.
\newblock {\em PeerJ Computer Science}, 2:e55, 2016.

\end{thebibliography}
\end{document}